\documentclass{vldb}
\usepackage{graphicx}
\usepackage{amsmath}
\usepackage{verbatim}
\usepackage{comment} 
\usepackage{algorithm}
\usepackage{algorithmic}
\usepackage{amsfonts}
\usepackage{amssymb}
\usepackage{subcaption}
\usepackage{multirow}
\usepackage{flushend}
\usepackage{caption}
\usepackage{xspace}
\usepackage{xcolor}
\usepackage{enumitem,url,placeins}

\newdef{theorem}{Theorem}
\newtheorem{problem}{Problem}
\newtheorem{lemma}{Lemma}
\newtheorem{definition}{Definition}

\begin{document}

\title{``The Whole Is Greater Than the Sum of Its Parts'' : Optimization in Collaborative Crowdsourcing}
\numberofauthors{1}
\author{
\alignauthor
 Habibur Rahman$^{\ddag}$, Senjuti Basu Roy$^{\dag}$, Saravanan Thirumuruganathan$^{\ddag}$ \\ ,Sihem Amer-Yahia$^{\diamond}$, Gautam Das$^{\ddag}$.  \\
\affaddr{
$^{\ddag}$UT Arlington,
$^{\dag}$UW Tacoma,
$^{\diamond}$ CNRS, LIG
}
{\email{
	senjutib@uw.edu, 
    \{habibur.rahman,saravanan.thirumuruganathan\}@mavs.uta.edu, 	
	sihem.amer-yahia@imag.fr,	
	gdas@uta.edu}
}
}

\maketitle

\begin{abstract}
In this work, we initiate the investigation of optimization opportunities in collaborative crowdsourcing. Many popular applications, such as collaborative document editing, sentence translation, or citizen science resort to this special form of human-based computing, where, crowd workers with appropriate skills and expertise are required to form {\em groups} to solve complex {\em tasks}. Central to any collaborative crowdsourcing process is the aspect of successful {\em collaboration} among the workers, which, for the first time, is {\em formalized and then optimized} in this work. Our formalism considers two main collaboration-related human factors, affinity and upper critical mass, appropriately adapted from organizational science and social theories.
Our contributions are (a) proposing a comprehensive model for collaborative crowdsourcing optimization, (b) rigorous theoretical analyses to understand the hardness of the proposed problems, (c) an array of efficient exact and approximation algorithms with provable theoretical guarantees. 
Finally, we present a {\em detailed set of experimental results} stemming from  two real-world collaborative crowdsourcing application using Amazon Mechanical Turk, as well as  conduct synthetic data analyses on scalability and qualitative aspects of our proposed algorithms. 
Our experimental results successfully demonstrate the efficacy of our proposed solutions.

\end{abstract}

\vspace{-0.15in}
\section{Introduction}
\label{introduction}
The synergistic effect of collaboration in group based activities is widely accepted in socio-psychological research and traditional team based activities~\cite{whole,whole1,andres2013team}. The very fact that the {\em collective yield of a group is higher than the sum of the contributions of the individuals} is often described as  
``the whole is greater than the sum of its parts''~\cite{whole,whole1}. Despite its immense potential, the transformative effect of ``collaboration'' remains largely unexplored in crowdsourcing~\cite{Kittur:2013:FCW:2441776.2441923} complex tasks (such as document editing, product design, sentence translation, citizen science), which are acknowledged as some of the most promising areas of next generation crowdsourcing. In this work, we investigate the optimization aspects of this specific form of human-based computation that involves people working in groups to solve complex problems that require collaboration and a variety of skills. We believe our work
is also the first to {\em formalize optimization in collaborative crowdsourcing.} 

The optimization goals of collaborative crowdsourcing are akin to that of its traditional micro-task based counterparts~\cite{bin1,bin2}  - quickly maximize the quality of the completed tasks, while minimizing cost, by {\em assigning appropriate tasks to appropriate workers}. However, the ``plurality optimization'' based solutions, typically designed for the micro-task based crowdsourcing are inadequate to optimize collaborative tasks, as the latter requires workers with certain skills to work in groups and ``build'' on each other's contributions for tasks that do not typically have ``binary'' answers.  Prior work in  collaborative crowdsourcing has proposed the importance of {\em human factors} to characterize workers, such as workers' skills and wages~\cite{DBLP:conf/dbcrowd/RoyLTAD13,DBLP:journals/corr/RoyLTAD14}. Additional human factors, such as {\em worker-worker affinity}~\cite{yantwo,kittur2011crowdforge}, is also acknowledged to quantify workers collaboration effectiveness. Similarly, social theories widely underscore the importance of {\em upper critical mass}~\cite{KM} for group collaboration, which is a constraint on the size of groups beyond which the collaboration effectiveness diminishes~\cite{KM,marwell1988social}. 
However, no further attempts have been made to formalize these variety of human factors in a principled manner to optimize the outcome of a collaborative crowdsourcing environment. 

{\bf Our first significant contribution lies in appropriately incorporating the interplay of these variety of complex human factors into a set of well-formulated optimization problems}. To achieve the aforementioned optimization goals, it is therefore essential to form, for each task, a group of workers who collectively hold skills required for the task, collectively cost less than the task's 
budget, and collaborate effectively.  Using the notions of affinity and upper critical mass, we
formalize the flat model of work coordination~\cite{hier} in collaborative
crowdsourcing as a graph with nodes representing workers and edges
labeled with pair-wise affinities. A group of workers is a clique in
the graph whose size does not surpass the critical mass imposed by a
task. A large clique (group) may further be partitioned into subgroups (each is a clique of smaller size satisfying critical mass) to complete a task because of the
task's magnitude. Each clique has an 
intra and an inter-affinity to measure respectively the level of cohesion that the
clique has internally and with other cliques. A clique with high
intra-affinity implies that its members collaborate well with one
another. Two cliques with a high inter-affinity between them implies
that these two groups of workers work well together.  Our optimization problem reduces to finding a clique that maximizes
intra-affinity, satisfies the skill threshold across multiple domains, satisfies the cost limit, and maximizes inter-affinity when partitioned into smaller cliques. We note that no existing work on team formation in social networks~\cite{Anagnostopoulos:2012:OTF:2187836.2187950,Lappas:2009:FTE:1557019.1557074} or collaborative crowdsourcing~\cite{Kittur:2013:FCW:2441776.2441923,yantwo,kittur2011crowdforge} has attempted similar formulations .

{\bf Our second endeavor is {\em computational}}. We show that solving the complex
optimization problem explained above is prohibitively expensive and
incurs very high machine latency. Such high latency is unacceptable for
a 
real-time 
crowdsourcing platform.  Therefore, we propose an alternative strategy {\tt Grp\&Splt} that decomposes the overall problem into two stages and is a natural alternative to our original problem formulation. Even though this staged formulation is also computationally intractable in the worst case, it allows us to design {\em instance optimal exact algorithms that work well in the average case, as well as efficient approximation algorithms with provable bounds}. In {\bf stage-1 (referred to as {\tt Grp})}, we first form a single group of workers by maximizing intra-affinity, while satisfying the skill and cost thresholds. In {\bf stage-2 (referred to as {\tt Splt})}, we decompose this large group into smaller subgroups, such that each satisfies the group size constraint (imposed by critical mass) and the inter-affinity across sub-groups is maximized. Despite being NP-hard~\cite{DBLP:books/fm/GareyJ79}, 
we propose an instance optimal {\em exact algorithm} {\tt OptGrp} 
and a novel {\em 2-approximation algorithm} {\tt ApprxGrp} for the stage-1 problem. 
Similarly, we prove the NP-hardness and propose a {\em 3-approximation} algorithm {\tt Min-Star-Partition} 
for a variant of the stage-2 problem. 

{\bf Finally, we conduct a comprehensive experimental study} with two different applications ({\em sentence translation} and {\em collaborative document editing}) using real world data from Amazon Mechanical Turk and present rigorous scalability and quality analyses using synthetic data.
Our experimental results demonstrate that our formalism is effective in aptly modeling the behavior of collaborative crowdsourcing and our proposed solutions are scalable.

In summary, this work makes the following contributions:

\begin{enumerate}[noitemsep]
\item {\em Formalism}: We initiate the investigation of optimization opportunities in collaborative crowdsourcing, identify and incorporate a variety of human factors in well formulated optimization problems.
\item {\em Algorithmic contributions}: We present comprehensive theoretical analysis of our problems and approaches. We analyze the computational complexity of our problems, and propose a principled staged solution. We propose exact instance optimal algorithms as well as  efficient approximation algorithms with provable approximation bounds.
\item {\em Experiments}: We present a comprehensive set of experimental results (two real applications as well as synthetic experiments) that demonstrate the effectiveness of our proposed solutions.
\end{enumerate}

The paper is organized as follows. Sections~\ref{sec:ex},~\ref{sec:dm},
and~\ref{sec:pbm} discuss a database application of collaborative crowdsourcing, our data model,
problem formalization, and initial solutions. Sections~\ref{stage1} and~\ref{stage2}
describe our theoretical analyses and proposed algorithmic solutions. Experiments are described  in~\ref{sec:exps}, related
work in Section~\ref{sec:related}, and conclusion are presented in
Section~\ref{sec:conclusion}. Additional results are presented in appendix.

\section{An Application}\label{sec:ex}
Sentence translation~\cite{chen2011building,yantwo,kittur2011crowdforge} is a frequently encountered application of collaborative crowdsourcing, where the objective is to use the crowd to build a translation database of sentences in different languages. Such databases later on serve as the ``training dataset'' for supervised machine learning algorithms for automated sentence translation purposes.

As a running example for this paper, consider a translation task $t$ designed for translating a English video clip to French. 
Typically, such translation tasks follows a 3-step process~\cite{yantwo,kittur2011crowdforge}: English speakers first translate the video in English, professional editors edit the translation, and finally workers with proficiency in both English and French translate English to French. 
Consequently, such task requires skills in $3$ different domains: English comprehension ($d_1$), English editing ($d_2$), and French Translation ability ($d_3$). 

In our optimization setting,
each task $t$ has a requirement of minimum skill per domain and maximum cost budget, and workers should  collaborate with each other (e.g., to correct each others' mistakes~\cite{yantwo}), and  
the collaboration effectiveness is quantified as the {\em affinity} of the group. Some aspects of our formulation has similarities with
team formation problems in social networks~\cite{Anagnostopoulos:2012:OTF:2187836.2187950}. The notion of affinity has been
identified in the related work on sentence translation tasks~\cite{yantwo,kittur2011crowdforge}, as well as team formation problems~\cite{Anagnostopoulos:2012:OTF:2187836.2187950}.

However, if the group is ``too large'', the effectiveness of collective actions diminishes~\cite{KM,marwell1988social} while undertaking the translation task, as an unwieldy group of workers fail to find effective assistance from their peers~\cite{yantwo,kittur2011crowdforge}.
Therefore, each task $t$ is associated with a corresponding upper critical mass constraint on the size of an effective group, i.e., a large group may need to be further decomposed into multiple subgroups in order to satisfy that constraint. A study of the importance of the upper critical mass constraint in the crowdsourcing context, as well as how to set its (application-specific) value, are important challenges that are best left to domain experts; however, we experimentally study this issue for certain applications such as sentence translation.

When this task arrives, imagine that there are $6$ workers $u_1, u_2,\ldots, u_6$ available in the crowdsourcing platform. Each worker has a skill value on each of the three skill domains described above, and a wage they expect. Additionally, the workers cohesiveness or affinity is also provided. These human factors of the workers are summarized in Tables~\ref{workerskilltable} 
and \ref{workeraffinitytable}, and the task requirements of $t$ (including thresholds on aggregated skill for each domain, total cost, and critical mass) are presented in Table~\ref{taskskilltable} and are further described in the next section.

The objective is to form a  ``highly cohesive'' group $\mathcal{G}$ of workers  that satisfies the lower bound of skill of the task and upper bound of cost requirements. Due to the upper critical mass constraint, $\mathcal{G}$ may further be decomposed into multiple subgroups. After that, each sub-group undertakes a subset of sentences to translate. 
Once all the subgroups finish their respective efforts, their contributions are merged. Therefore, both the overall group and its subgroups must be cohesive.
Incorporation of upper critical mass makes our problem significantly different from the body of prior works~\cite{Anagnostopoulos:2012:OTF:2187836.2187950}, as we may have to create a group further decomposed into mutiple subgroups, instead of a single group.

\begin{table}
\centering  
    \begin{tabular}{| l | l | l | l | l | l | l |}
    \hline
     & $\mathbf{u_1}$ & $\mathbf{u_2}$ & $\mathbf{u_3}$ & $\mathbf{u_4}$ & $\mathbf{u_5}$ & $\mathbf{u_6}$ \\ 
    \textbf{$d_1$} & 0.66 & 1.0 & 0.53 & 0.0 & 0.13 & 0.0 \\
    \textbf{$d_2$} & 0.0 & 0.0 & 0.66 & 0.73 & 0.66 & 0.13 \\
    \textbf{$d_3$} & 0.0 & 0.33 & 0.53 & 0.0 & 0.8 & 0.93\\
    \textbf{Wage} & 0.4 & 0.3 & 0.7 & 0.8 & 0.5 & 0.8\\ 
	\hline
  \end{tabular}
\vspace{-0.05in}   
   \caption{\small Workers skill and wage table}
  \label{workerskilltable}
\end{table}
  \vspace{-0.05in}

  \vspace{-0.05in}
\begin{table}
\centering
    \begin{tabular}{| l | l | l | l | l | l | l |}
    \hline
     & $\mathbf{u_1}$ & $\mathbf{u_2}$ & $\mathbf{u_3}$ & $\mathbf{u_4}$ & $\mathbf{u_5}$ & $\mathbf{u_6}$ \\ 
    $\mathbf{u_1}$ & 0.0 & 1.0 & 0.66 & 0.66 & 0.85 & 0.66 \\
    $\mathbf{u_2}$ & 1.0 & 0.0 & 0.66 & 0.85 & 0.66 & 0.85 \\
    $\mathbf{u_3}$ & 0.66 & 0.66 & 0.0 & 0.4 & 0.66 & 0.40\\
    $\mathbf{u_4}$ & 0.66 & 0.85 & 0.4 & 0.0 & 0.4 & 0.0\\ 
    $\mathbf{u_5}$ & 0.85 & 0.66 & 0.66 & 0.4 & 0.0 & 0.4\\ 
    $\mathbf{u_6}$ & 0.66 & 0.85 & 0.4 & 0.0 & 0.4 & 0.0\\ 
	\hline
  \end{tabular}
  \vspace{-0.05in}
   \caption{\small Workers Distance Matrix}
  \label{workeraffinitytable}
\end{table}

\begin{table}
\centering
\begin{tabular}{| p{2em} | p{2em} | p{2em} | p{2em} | p{2em}|}
    \hline
    $Q_1$ & $Q_2$ & $Q_3$ & C & K \\ 
	\hline    
    1.8 & 1.4 & 1.66 & 3.0 & 3 \\
    \hline
  \end{tabular}
  \vspace{-0.05in}
  \caption{\small Task Description}\label{taskskilltable}
\end{table}

\newpage
\section{Data Model}
\label{sec:dm}

We introduce our data model and preliminaries that will serve as a
basis for our problem definition. 

\vspace{-0.1in}
\subsection{Preliminaries}

{\bf Domains:} We are given a set of domains $D=\{d_1,d_2,\ldots,d_{m}\}$  denoting knowledge topics.
Using the running example in Section~\ref{sec:ex}, there are $3$ different domains - English comprehension ($d_1$), English editing ($d_2$), and French Translation ability($d_3$).

{\bf Workers:} We assume a set $\mathcal{U}=\{u_1,u_2,\ldots,u_{n}\}$
of $n$ workers available in the crowdsourcing platform. The example in Section~\ref{sec:ex} describes a crowdsourcing platform with $6$ workers.

\smallskip {\bf Worker Group:} A worker group $\mathcal{G}$ consists of a subset of workers from $\mathcal{U}$ 
i.e. $\mathcal{G} \subseteq \mathcal{U}$.

\smallskip  {\bf Skills:} A skill is the knowledge on a particular skill domain in
$D$, quantified in a continuous $[0,1]$ scale. It is associated with
workers and tasks. 
The skill of a worker represents the worker's expertise/ability on a topic. The skill of a topic represents the minimum knowledge requirement/quality for that task.
A value of
$0$ for a skill reflects no expertise of a worker for that skill. For
a task, $0$ reflects no requirement for that skill.

How to learn the skill of the workers is an important and independent research problem in its own merit. Most related work has relied on learning skill of the workers from ``gold-standard'' or benchmark datasets using pre-qualification tests ~\cite{Downs:2010:YPG:1753326.1753688,Josang:2007:STR:1225318.1225716}. As we describe in Section~\ref{sec:us} in detail, we also learn the skill of the workers by designing pre-qualification tests using benchmark datasets.

\smallskip  {\bf Collaborative Tasks:} A collaborative task $t$ has the following
characteristics - a minimum knowledge threshold $Q_{i}$ per domain
$d_i$ in $D$, a maximum cost budget $C$ for hiring workers to achieve
$t$, and an upper critical mass $K$, denoting the maximum number of workers who can
effectively collaborate  inside a group to complete $t$. 
Specifically, $t$ is characterized by a vector, $\langle Q_{1}, Q_{2}, \ldots, Q_{m}, C, K
\rangle$, of length $m+2$. For the example in Section~\ref{sec:ex}, there are $3$ domains ($m=3$) and their respective skill requirements, its cost $C$, and critical mass $K$ of the task is described in Table~\ref{taskskilltable}. 
A task is considered complete if it attains its skill requirement over all domains and satisfies all the constraints.


\subsection{Human Factors}
A worker is described by a set of human
factors.  We consider two types of factors - factors that describe
individual worker's characteristics and factors that characterize an
individual's ability to work with fellow workers. Our contribution is in appropriately adapting these factors in collaborative crowdsourcing from multi-disciplinary prior works such as team formation~\cite{Anagnostopoulos:2012:OTF:2187836.2187950,Lappas:2009:FTE:1557019.1557074} and psychology research~\cite{KM,marwell1988social}. 

\subsubsection{Individual Human Factors: Skill and Wage} 
Individual workers in a crowdsourcing environment are characterized by
their skill and wage.

\smallskip \textbf{Skill}: For each knowledge domain $d_i$, $u_{d_i} \in [0,1]$
is the expertise level of worker $u$ in $d_i$. Skill expertise
reflects the quality that the worker's contribution has on a
task accomplished by that worker.

\smallskip \textbf{Wage}: $w_u \in [0,1]$ is the minimum amount of compensation
for which a worker $u$ is willing to complete a task. We choose a
simple model where a worker specifies a single wage value independent
of the task at-hand.

Table~\ref{workerskilltable} presents the respective skill of the $6$ workers in $3$ different domains and their individual wages for the running example.

\subsubsection{Group-based Human Factors: Affinities}
Although related work in collaborative crowdsourcing acknowledges the importance of workers' affinity to enable effective collaboration~\cite{yantwo,kittur2011crowdforge}, there is no attempt to formalize the notion any further. A worker's effectiveness in collaborating with her fellow workers is measured as {\em affinity}. 
We adopt an affinity model similar to group formation problems in social networks~\cite{lappas2009finding,Anagnostopoulos:2012:OTF:2187836.2187950}, where the atomic unit of affinity is {\em pairwise}, i.e., a measure of cohesiveness between every pair of workers. After that, we 
propose different ways to capture {\em  intra-group} and {\em inter-group}
affinities.

{\bf Pairwise affinity:} The affinity between two workers $u_i$ and
$u_j$, $\mathit{aff}(u_i,u_j)$, can be calculated by capturing the {\em similarity} between workers using simple socio-demographic attributes, such as region, age, gender, as done in previous work~\cite{yantwo}, as well as more complex psychological characteristics~\cite{MBTI}. For our purpose, we normalize pairwise
affinity values to fit in $[0,1]$ and use a notion of worker-worker {\em distance}
instead, i.e., where $\mathit{dist}(u_i,u_j) = 1 - \mathit{aff}(u_i,u_j)$. Thus a smaller distance between
workers ensures a better collaboration. Table~\ref{workeraffinitytable} presents the pair-wise distance of all $6$ workers for running example in Section~\ref{sec:ex}. As will be clear later, the notion of distance rathey than affinity enables the design of better algorithms for our purposes.

{\bf Intra-group affinity:} For a group $\mathcal{G}$, its intra-group affinity
measures the collaboration effectiveness among the workers in $\mathcal{G}$. Here
again we use distance and compute intra-group distance in one of two
 natural ways: computing the
diameter of $\mathcal{G}$ as the largest distance between any two workers in
$\mathcal{G}$, or aggregating all-pair worker distances in $\mathcal{G}$:
\begin{align*}
\mathit{DiaDist}(\mathcal{G}) &= Max_{\forall u_i,u_j \in \mathcal{G}} \mathit{dist}(u_i,u_j)\\ 
\mathit{SumDist}(\mathcal{G}) &= \Sigma_{\forall u_i,u_j \in \mathcal{G}} \mathit{dist}(u_i,u_j)
\end{align*}

For both definitions, smaller value is better. 

{\bf Inter-group affinity:} When a group violates the upper critical mass
constraint~\cite{KM}, it needs to be decomposed into multiple smaller ones.
The resulting subgroups need to work together to achieve the task. Given two subgroups $G_1, G_2$ split from a
large group $\mathcal{G}$, their 
collaboration effectiveness is captured by computing their inter-group
affinities. Here again, we use distance instead of affinity. More concretely, the
inter-group distance is defined in one of two natural ways: either the largest distance between any two workers across the sub-groups, or the aggregation of all pair-wise workers
distances across subgroups: 
\begin{align*}
\mathit{DiaInterDist}(G_1,G_2) &= Max_{\forall u_i \in G_1, u_j \in G_2} \mathit{dist}(u_i,u_j)\\
\mathit{SumInterDist}(G_1,G_2) &= \Sigma_{\forall u_i \in G_1, u_j \in G_2} \mathit{dist}(u_i,u_j)
\end{align*}
This can be generalized to more than two subgroups: if there are $x$ subgroups, overall inter-group affinity is the summation of inter-group affinity for all possible pairs (${^x}C_2)$.


\section{Optimization}\label{sec:pbm}

{\bf Problem Settings:} For each collaborative task, we intend to form {\em the most appropriate group of workers} from the available worker pool. A collaborative crowdsourcing task has skill requirements in multiple domains and a cost budget, which is similar to the requirements of collaborative tasks in team formation problems~\cite{lappas2009finding}. Then, we adapt the ``flat-coordination'' models of worker interactions, which is considered  important in prior works in team formation~\cite{Anagnostopoulos:2012:OTF:2187836.2187950} as the ``coordination cost'', or in collaborative crowdsourcing~\cite{yantwo} itself, as `the `turker-turker'' affinity model.  However, unlike previous work, we attempt to fully explore the potential of ``group synergy''~\cite{surowiecki2004wisdom} and how it yields the maximum qualitative effects in group based efforts by maximizing affinity among the workers (or minimizing distance). Finally, we intend to investigate the effect of upper critical mass in the context of collaborative crowdsourcing as a constraint on group size, beyond which the group must be decomposed into multiple subgroups that are cohesive inside and across. Indeed, our objective function is designed to form a group (or further decomposed into a set of subgroups) to undertake a specific task that achieves the highest qualitative effect, while satisfying the cost constraint.


(1) {\em Qualitative effect of a group}: Intuitively, the overall qualitative effect  of a formed group to undertake a specific task is a function of the skill of the workers and their collaboration effectiveness. Learning this function itself is challenging, as it requires access to adequate training data and domain knowledge. In our initial effort, we therefore make a reasonable simplification, where we seek to maximize group affinity and pose quality as a hard constraint\footnote{\small Notice that posing affinity as a constraint does not fully exploit the effect of ``group synergy''.}. Existing literature (indicatively~\cite{surowiecki2004wisdom}) informs us that aggregation is a mechanism that turns private judgments (in our case individual workers' contributions) into a collective decision (in our case the final translated sentences), and is one of the four pillars for the wisdom of the crowds. For complex tasks like sentence translation or document editing, there is no widely accepted mathematical function of aggregation. We choose sum to aggregate the skill of the workers that must satisfy the lower bound of the quality of the task. This simplest and yet most intuitive functions for transforming individual contributions into a collective result  has been adopted in many previous works~\cite{Anagnostopoulos:2012:OTF:2187836.2187950,lappas2009finding,gajewar2012multi}. Moreover, this simpler function allows us to design efficient algorithms. Exploring other complex functions (e.g., multiplicative function) or learning them  is deferred to future work.

(2){\em Upper critical mass}: Sociological theories widely support the notion of ``critical mass''\cite{KM,marwell1988social} by reasoning that large groups are less likely to support collective action. However, whether the effect of ``critical mass'' should be imposed as a hard constraint, or it should have more of a gradual ``diminishing return'' effect, is itself a research question. For simplicity, we consider upper critical mass as a hard constraint and evaluate its effectiveness empirically for different values. Exploring more sophisticated function to capture critical mass is deferred to future work.\\

\begin{problem}
{\bf AffAware-Crowd:} Given a collaborative task $t$, the objective is to {\em form a worker group} $\mathcal{G}$, further partitioned into a set of $x$ subgroups $G_1,G_2,....G_x$ (if needed) for the task $t$ that minimizes the aggregated intra-distance of the workers, as well as the aggregated inter-distance across the subgroups of $\mathcal{G}$, and $\mathcal{G}$ must satisfy the skill and cost thresholds of $t$, where each subgroup $G_i$ must satisfy  the upper critical mass constraint of $t$. Of course, if the group $\mathcal{G}$ itself satisfies the critical mass constraint, no further partitioning in $\mathcal{G}$ is needed, giving rise to a single worker group. As explained above, quality of a task is defined as an aggregation (sum) of the skills of the workers~\cite{Anagnostopoulos:2012:OTF:2187836.2187950,lappas2009finding}. Similarly, cost of the task is the additive wage of all the workers in $\mathcal{G}$. 
\end{problem}


\subsection{Optimization Models}
Given the high-level definition above, we propose multiple optimization objective functions based on different inter- and intra-distance measures defined in Section~\ref{sec:dm}. 

For a group $\mathcal{G}$, we calculate intra-distance in one of the two possible ways: $\mathit{DiaDist()},\mathit{SumDist()}$. If $\mathcal{G}$ is further partitioned to satisfy the upper critical mass constraint, then we also want to enable strong collaboration across the subgroups by minimizing inter-distance. For the latter, inter-distance is calculated using one of $\mathit{DiaInterDist()}, \mathit{SumInterDist()}$. 
Even though there may be many complex formulations to combine these two factors, in our initial effort our overall objective function is a simple sum of these two factors that we wish to minimize. This gives rise to $4$ possible optimization objectives.
\begin{itemize}
\item {\bf $\mathit{DiaDist()},\mathit{DiaInterDist()}$:} 
\begin{align*} 
& \text{Minimize } \{\mathit{DiaDist}(\mathcal{G}) \quad + & \\ 
& 
\qquad \qquad 
Max\{
\forall G_i,G_j \in \mathcal{G} \quad
\mathit{DiaInterDist}(G_i,G_j)\} \} & \\ 
\end{align*} 

\item {\bf $\mathit{SumDist()},\mathit{DiaInterDist()}$:} 
\begin{align*} 
& \text{Minimize } \{\mathit{SumDist}(\mathcal{G}) \quad + & \\ 
& \qquad \qquad 
Max\{
\forall G_i,G_j \in \mathcal{G} \quad 
\mathit{DiaInterDist}(G_i,G_j)\} \} & \\ 
\end{align*}

\item {\bf $\mathit{DiaDist()},\mathit{SumInterDist()}$:}
\begin{align*} 
\text{Minimize } \{\mathit{DiaDist}(\mathcal{G}) + \sum_{\forall_{G_i,G_j \in \mathcal{G}}}
\! \! \! \!  
\mathit{SumInterDist}(G_i,G_j)\}
\end{align*}

\item {\bf $\mathit{SumDist()},\mathit{SumInterDist()}$:} 
\begin{align*} 
\text{Minimize }  \{\mathit{SumDist}(\mathcal{G}) + \sum_{\forall_{G_i,G_j \in \mathcal{G}}} 
\! \! \! \!  
\mathit{SumInterDist}(G_i,G_j)\}
\end{align*}
\end{itemize}

where, each of these objective function has to satisfy the  following three constraints on skill, cost, and critical mass respectively, as described below:
\begin{align*} 
\Sigma_{\forall_{u_i \in \mathcal{G}}} u_{d_i} &\geq Q_i \quad \forall_{d_i} \\
\Sigma_{\forall_{u \in \mathcal{G}}} w_u  &\leq C \\
 |G_i| &\leq K \quad \forall i=\{1,2,\ldots,x\}
\end{align*} 


For brevity, the rest of our discussion only considers  $\mathit{DiaDist()}$ on intra-distance and $\mathit{SumInterDist()}$ on inter-distance. We refer to this variant of the problem as {\tt AffAware-Crowd}. We note that our proposed optimal solution in Section~\ref{sec:pbm} could be easily extended to other combinations as well.

 



\vspace{-0.1in}

\begin{theorem}\label{th1}
Problem {\tt AffAware-Crowd} is NP-hard~\cite{DBLP:books/fm/GareyJ79}.
\end{theorem}

The detailed proof is provided in the appendix inside Section~\ref{proofs}.

\subsection{Algorithms for AffAware-Crowd}
Our optimization problem attempts to appropriately capture the complex interplay among various important factors.  The proof of Theorem~\ref{th1} in Section~\ref{proofs} in the appendix shows that even the simplest variant of the optimization problem is NP-hard. Despite the computational hardness, we attempt to stay as principled as possible in our technical contributions and algorithms design. Towards this end, we propose two alternative directions: (a) We investigate an integer linear programming (ILP)~\cite{ip} formulation to optimally solve our original overarching optimization problem. We note that even translating the problem to an ILP is non-trivial, because the subgroups inside the large group are also unknown and are determined by the solution. ( b) Since ILP is prohibitively expensive (as our experimental results show), we propose an alternative strategy that is natural to our original formulation, referred to as {\tt Grp\&Splt}. {\tt Grp\&Splt} decomposes the original problem into two phases: in the {\tt Grp} phase, a single group is formed that satisfies the skill and cost threshold, but ignores the upper critical mass constraint. Then, in the {\tt Splt} phase, we partition this large group into a set of subgroups, each satisfying the upper critical mass constraint, such that the sum of all pair inter-distance is minimized. Note that, for many tasks, the {\tt Grp} stage itself may be adequate, and we may never need to execute {\tt Splt}. We propose a series of efficient polynomial time approximation algorithms for each phase, each of which has a provable approximation factor. Of course, this staged solution combined together may not have any theoretical guarantees for our original problem formulation. However, our experimental results demonstrate that this formulation is efficient, as well as adequately effective.


\subsubsection{ILP for AffAware-Crowd}\label{ip}
\vspace{-0.2in}
\begin{equation}\label{ilp}
\begin{aligned}
& {\text{minimize}} & & \mathcal{D} = \mathrm{Max} \{ e_{i, i'} \times dist(u_i, u_{i'}) \} \quad +  \\ 
& & & \qquad \sum_{\forall G_i, G_j \in \mathcal{G}} \quad \sum_{ \forall u_i \in G_i, u_j \in G_j} e_{i,j} dist(u_i, u_j) \\
& \text{subject to} & & & \\
& & & \sum_{i=1}^{n} \sum_{j=1}^{x} u_{(i,G_j)} \times u^i_{d_l} \geq Q_l 
    \qquad 
    \forall l \in [1,m] \\
& & & \sum_{i=1}^{n} \sum_{j=1}^{x} u_{(i,G_j)} \times w_{u}^i  \leq C \\
& & & \sum_{i=1}^{n} u_{(i, G_j)} \leq K 
    \qquad \qquad \quad \quad 
    \forall j \in [1,x] \\
& & & \sum_{j=1}^{x} u_{(i,G_j)} \leq 1 
    \qquad \qquad \qquad \: \: 
    \forall i \in [1,n] \\
& & & e_{i,i'} = 
\begin{cases}
1       & \exists j \in [1,x] \: \& \: u_{(i,G_j)} = 1 \: \& \: u_{(i',G_j)} = 1\\
0       & \text{otherwise} \\
\end{cases} \\
& & & x \in \{0,1,\ldots,n\}  \\ 
& & & u_{(i,G_j)} \in \{0,1\} 
    \qquad \qquad \quad \: 
    \forall i \in [1,n], \forall j \in [1,x] \\
\end{aligned}
\end{equation}

We discuss the ILP next as shown in Equation~\ref{ilp}. Let $e_{(i,i')}$ denote a boolean decision variable of whether a user pair $u_i$ and $u_i'$ would  belong to same sub-group in group $\mathcal{G}$ or not. Also, imagine that a total of $x$ groups ($G_1,G_2, \ldots, G_x$) would be formed for task $t$, where $1 \leq x \leq n$ (i.e., at least the subgroup is $\mathcal{G}$ itself, or at most $n$ singleton subgroups could be formed). Then, which subgroup the worker pair should be assigned must also be determined, where the number of subgroups is unknown in the first place. Note that translating the problem to an ILP is {\em non-trivial and challenging}, as the formulation deliberately makes the problem linear by translating each worker-pair as an atomic decision variable (as opposed to a single worker) in the formulation, and it also returns the subgroup where each pair should belong to. Once the ILP is formalized, we use a general-purpose solver to solve it. Although the {\em Max} operator in the objective function (expresses $\mathit{DiaDist()}$) must be translated appropriately further in the actual ILP implementation, in our formalism below, we preserve that abstraction for simplicity.

The objective function returns a group of subgroups that minimizes $\mathit{DiaDist}(\mathcal{G}) + \Sigma_{\forall_{G_i,G_j}}\mathit{SumInterDist}(G_i,G_j)$. The first three constraints ensure the skill, cost and upper critical mass thresholds, whereas the last four constraints ensure the disjointedness of the group and the integrality constraints on different Boolean decision variables.

When run on the example in Section~\ref{sec:ex}, the ILP generates the optimal solution and creates group $\mathcal{G} = \{u_1,u_2,u_3,u_4,u_6\}$ with two subgroups, $G_1 = \{u_1,u_2,u_4\}$, and  
$G_2 = \{u_3,u_6\}$. The distance value of the optimization objective is $4.23$, which equals to $\mathit{DiaDist}(\mathcal{G}) + \mathit{InterDist}(G_1,G_2)$.

\subsubsection{Grp\&Splt : A Staged Approach}\label{tier}
Our proposed alternative strategy {\tt Grp\&Splt} works as follows: in the {\bf \tt Grp} stage, we attempt to form {\em a single worker group that minimizes $\mathit{DiaDist}(\mathcal{G})$, while satisfying  the skill and cost constraints (and ignoring the upper critical mass constraint)}. Note that this may result in a large group, violating the upper critical mass constraints. Therefore, in the {\bf \tt Splt} phase, we {\em partition this big group into multiple smaller sub-groups, each satisfying the upper critical mass constraint in such a way that the aggregated inter-distance between all pair of groups $\Sigma_{\forall_{Gi,Gj}}\mathit{SumInterDist}(G_i,G_j)$ is minimized}. As mentioned earlier, there are three primary reasons for taking this alternative route: (a) In many cases we may not even need to execute {\tt Splt}, because the solo group formed in {\tt Grp}  phase abides by the upper critical mass constraint leading to the solution of the original problem. (b) The original complex ILP is prohibitively expensive. Our experimental results demonstrate that the original ILP does not converge in hours for more than $20$ workers. (c) Most importantly, {\tt Grp\&Splt} allows us to design  {\em efficient approximation algorithms with constant approximation factors} as well as instance optimal exact algorithms that work well in practice, as long as the distance between the workers satisfies the {\em metric property} (triangle inequality in particular) \cite{fd1,fd2}. We underscore that the triangle inequality assumption is not an overstretch, rather many natural distance measures (Euclidean distance, Jaccard Distance) are metric and several other similarity measures, such as Cosine Similarity, Pearson and Spearman Correlations could be transformed to metric distance~\cite{metric}. Furthermore, this assumption has been extensively used in distance computation in the related literature~\cite{Anagnostopoulos:2010:PUF:1871437.1871515,Anagnostopoulos:2012:OTF:2187836.2187950}. Without metric property assumptions, the problems remain largely inapproximable~\cite{fd2}.

\section{Enforcing Skill \& Cost : GRP }\label{stage1}	
In this section, we first formalize our proposed approach in {\tt Grp} phase, discuss hardness results, and propose algorithms with theoretical guarantees. Recall that our objective is to form a single group $\mathcal{G}$ of workers that are cohesive (the diameter of that group is minimized), while satisfying the skill and the cost constraint.

\begin{definition}
{\tt Grp}: Given a task $t$, form a single group $\mathcal{G}$ of workers that minimizes $\mathit{DiaDist}(\mathcal{G})$, while satisfying the skill and cost constraints, i.e., $\Sigma_{\forall u \in \mathcal{G}} u_{d_i} \geq Q_i, \forall_{d_i}$, \&  $\Sigma_{\forall u \in \mathcal{G}} w_u \leq C$.
\end{definition}

\begin{theorem}
Problem {\tt Grp} is NP-hard.
\end{theorem}

The detailed proof is discussed in Section~\ref{proofs} in appendix.

{\bf Proposed Algorithms for {\tt Grp}:}
We discuss two algorithms at length - a) {\tt OptGrp} is an instance optimal algorithm.  b) {\tt ApprxGrp} algorithm has a {\em $2$-approximation factor}, as long as the distance satisfies the triangle inequality property. Of course, an additional optimal algorithm is the ILP formulation itself (referred to as {\tt ILPGrp} in experiments), which could be easily adapted from Section~\ref{sec:pbm}. Both {\tt OptGrp} and {\tt ApprxGrp} invoke a subroutine inside, referred to as {\tt GrpCandidateSet()}. We describe a general framework for this subroutine next.
%
%

\subsection{Subroutine GrpCandidateSet()}
Input to this subroutine is a set of $n$ workers and a task $t$ (in particular the skill and the cost constraints of $t$) and the output is a {\em worker group} that satisfies the skill and cost constraints. Notice that, if done naively, this computation takes $2^n$ time. However, Subroutine {\tt GrpCandidateSet()} uses effective pruning strategy to avoid unnecessary computations that is likely to terminate much faster. It {\em computes a binary tree representing the possible search space} considering the nodes in an arbitrary order, each node in the tree is a worker $u$ and has two possible edges ($1/0$, respectively stands for whether $u$ is included in the group or not). A root-to-leaf path in that tree represents a {\em worker group}.

\begin{figure}[h]
\centering
\includegraphics[width=3.0in]{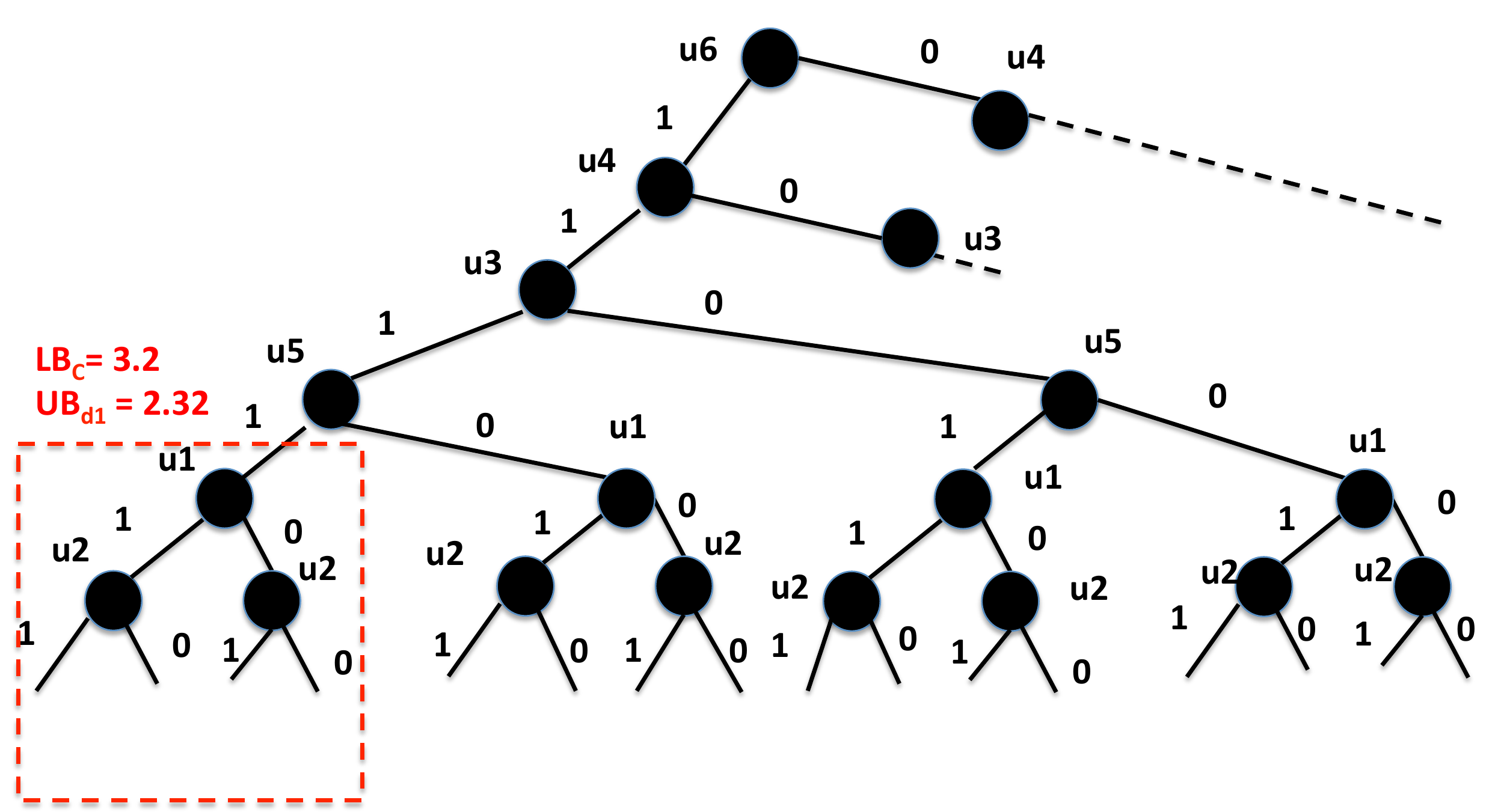}
\vspace{-0.1in}
\caption{\small \label{tree} A partially constructed tree of GrpCandidateSet() using the example in Section~\ref{sec:ex}. At node $u_1=1$, $LB_C$ = $w_{u_6}+w_{u_4}+w_{u_3}+w_{u_5} + w_{u_1} = 3.2$ and $UB_{d_1} = u^6_{d_1}+u^4_{d_1}+u^3_{d_1}+ u^5_{d_1}+u^1_{d_1}+ u^2_{d_1} = 2.32$. The entire subtree is pruned, since $LB_C (3.2) > C$.}
\end{figure}

At   a given node $u$, it makes two estimated bound computation : a) it computes the lower bound of cost ($LB_C$) of that path (from the root upto that node), b) it computes the upper bound of skill of that path ($UB_{d_i}$) for each domain. It compares $LB_C$ with $C$ and compares $UB_{d_i}$ with $Q_i, \forall d_i$. If $LB_C > C$ or $UB_{d_i} < Q_i$ for any of the domains, that branch is fully pruned out. Otherwise, it continues the computation. Figure~\ref{tree} has further details.

 {\tt ApprxGrp()} uses this subroutine to find the first valid answer, whereas, Algorithm {\tt OptGrp()} uses it to return all valid answers.

\subsection{Further Search Space Optimization}\label{cons}
When the skill and cost of the workers are arbitrary, a keen reader may notice that Subroutine {\tt GrpCandidateSet()} may still have to explore $2^n$ potential groups at the worst case. Instead, if we have only a constant number of costs and arbitrary skills, or a constant number of skill values and any arbitrary number of costs, interestingly, the search space becomes polynomial. Of course, the search space is polynomial when both are constants.


We describe the constant cost idea further. Instead of any arbitrary wage of the workers, we now can discretize workers wage apriori and create a constant number of $k$ different buckets of wages (a worker belongs to one of these buckets) and build the search tree based on that.  When there are $m$ knowledge domains, this gives rise to a total of $mk$ buckets. For our running example in Section~\ref{sec:ex}, for simplicity if we consider only one skill ($d_1$), this would mean that we discretize all $6$ different wages in $k$ (let us assume $k=2$) buckets.  Of course, depending on the granularity of the buckets this would introduce some approximation in the algorithm as now the workers actual wage would be replaced by a number which may be lesser or greater than the actual one. However, such a discretization may be realistic, since many crowdsourcing platforms, such as AMT, allow only one cost per task. 

\begin{figure}[h]
\includegraphics[width=2.5in]{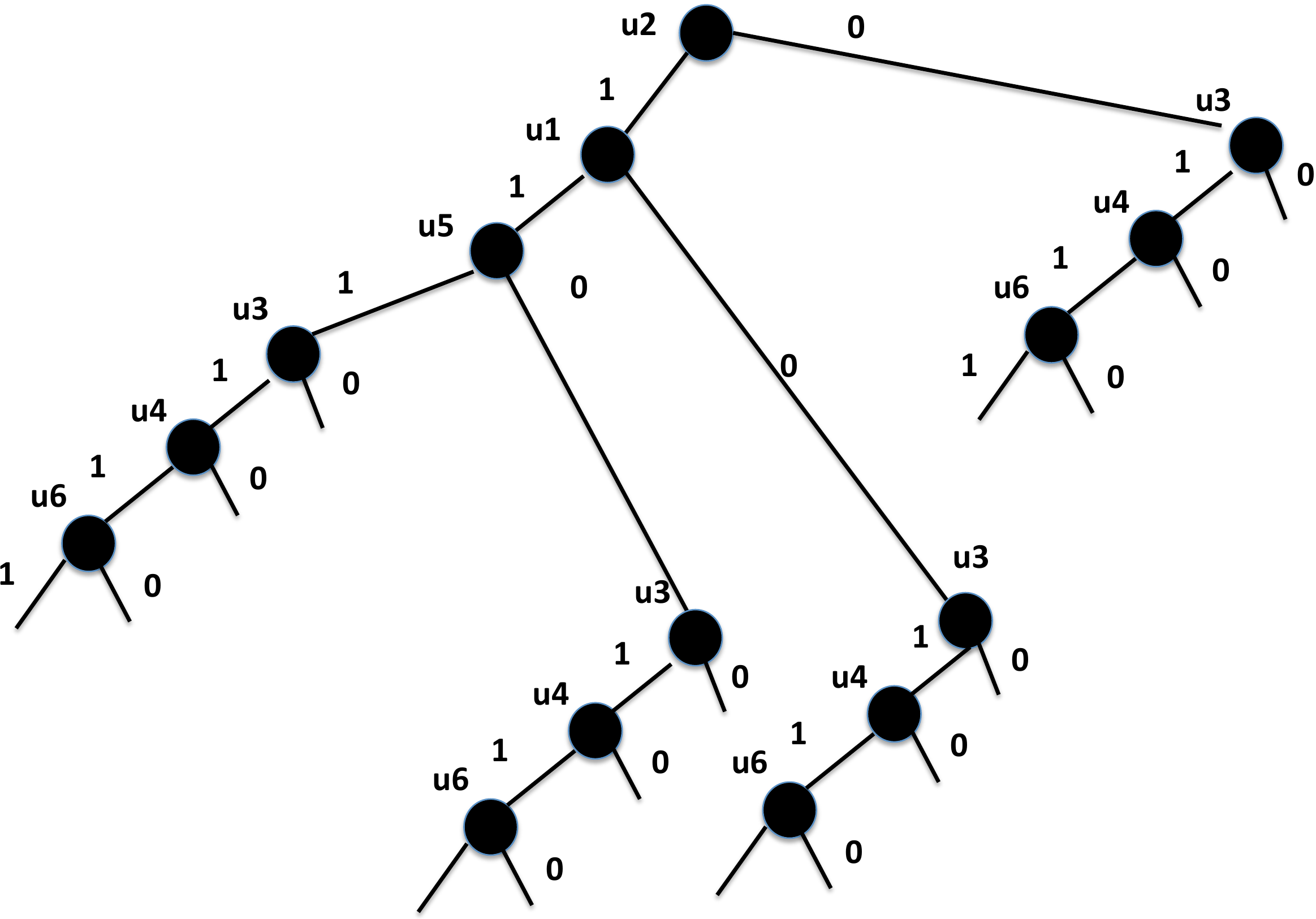}
\vspace{-0.1in}
\caption{\small \label{tree2} Possible search space using the example in Section~\ref{sec:ex}, after the cost of the workers are discretized into $k=2$ buckets, considering only one skill $d_1$. The tree is constructed in descending order of skill of the workers per bucket. For bucket $1$, if the most skilled worker {\em $u_2$ is not selected, the other two workers ($u_1,u_5$) will never be selected}.}
\end{figure}

For our running example, let us assume, bucket 1 represents wage $0.5$ and below, bucket 2 represents wage between $0.5$ and $0.8$. Therefore, now workers $u_3,u_4,u_6$ will be part of bucket $2$ and the three remaining workers will be part of bucket $1$.  After this, one may notice that the tree will neither be balanced nor exponential. Now, for a given bucket, the possible ways of worker selection is polynomial (they will always be selected from most skilled ones to the least skilled ones), making the overall search space polynomial for a constant number of buckets. In fact, as opposed to $2^6$ possible branches, this modified tree can only have $(3+1) \times (3+1)$ possible branches. Figure~\ref{tree2} describes the idea further.

Once this tree is constructed, our previous pruning algorithm {\tt GrpCandidateSet()} could be applied to enable further efficiency. 
\vspace{-0.1in}
\vspace{-0.1in}


\vspace{-0.1in}
\subsection{Approximation Algorithm ApprxGrp}\label{apx}
A popular variant of facility dispersion problem~\cite{fd1,fd2} attempts to discover a set of nodes (that host the facilities) that are as far as possible, whereas, compact location problems~\cite{Krumke96compactlocation} attempt to minimize the diameter. For us, the workers are the nodes, and {\tt Grp} attempts to find a worker group that minimizes the diameter, while satisfying the multiple skills and a single cost constraint. We propose a $2$-approximation algorithm for {\tt Grp}, that is not studied before.

\begin{figure}[h]
\centering
\includegraphics[width=1.0in]{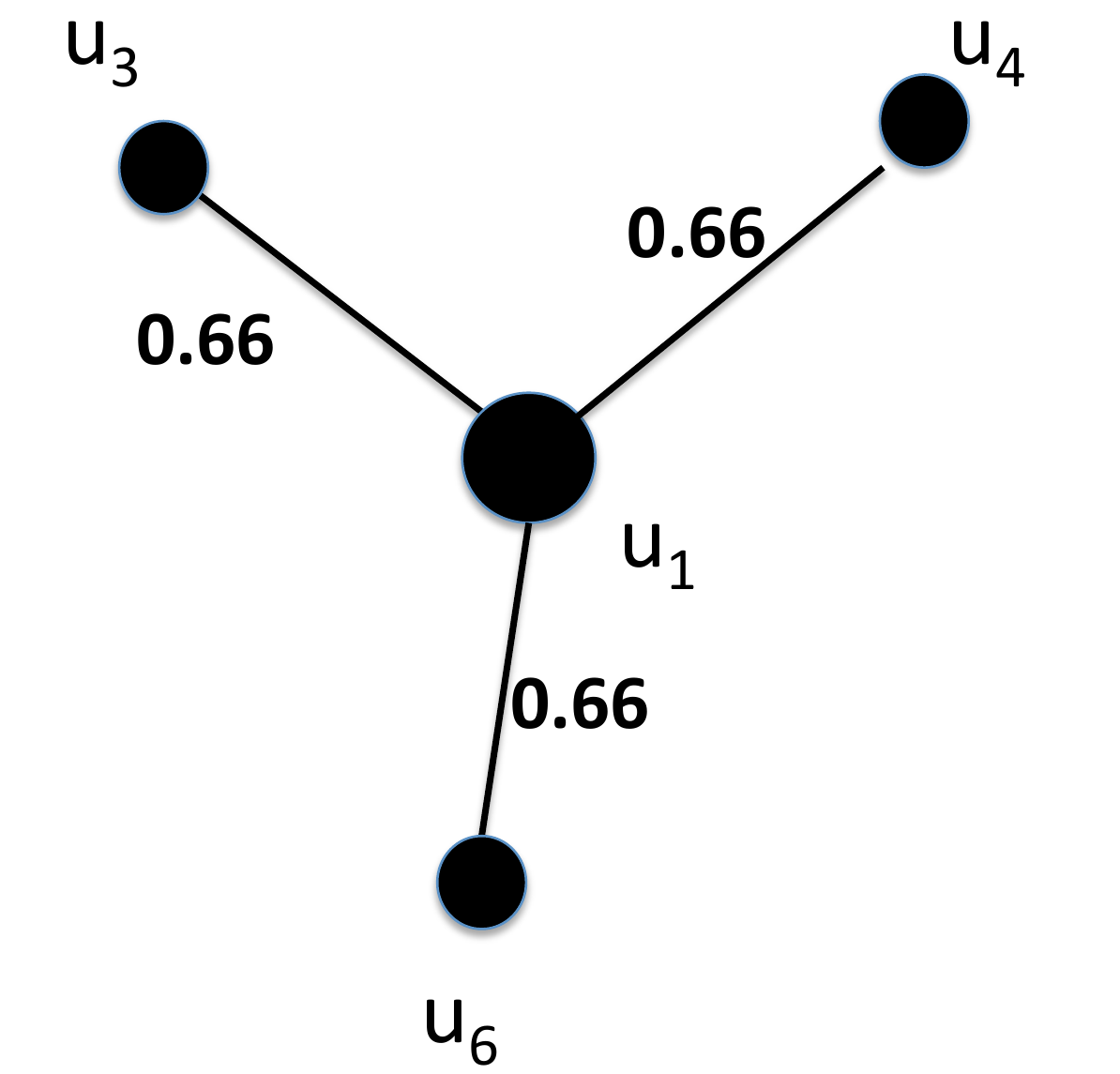}
\caption{\label{stargraph} An instantiation of ${\tt GrpDia(0.66)}$  using the example in Section~\ref{sec:ex}. A star graph centered $u_1$ is formed.}
\end{figure}

Algorithm {\tt ApprxGrp} works as follows: 
The main algorithm considers a sorted (ascending) list $\mathcal{L}$ of distance values (this list represents all unique distances between the available worker pairs in the platform) and performs a binary search over that list. First, it calls a subroutine ({\tt GrpDia()}) with a distance value $\alpha$ that can run at the most $n$ times. Inside the subroutine, it considers worker $u_i$ in the $i$-th iteration to retrieve a {\em star graph}\footnote{\small{Star graph is a tree on $v$ nodes with one node having degree $v-1$ and other $v-1$ nodes with degree 1.}}
centered around $u_i$ that satisfies the distance $\alpha$. The nodes of the star are the workers and the edges are the distances between each worker pair, such that no edge in that retrieved graph has an edge $>\alpha$. One such star graph is shown in Figure~\ref{stargraph}.

Next, given a star graph with a set of workers $\mathcal{U}'$,  {\tt GrpDia} invokes {\tt GrpCandidateSet($\mathcal{U}',t$)} to select a subset of workers (if there is one) from $\mathcal{U}'$, who together satisfy the skill and cost thresholds. {\tt GrpCandidateSet} constructs the tree in the best-first-search manner and terminates when the first valid solution is found, or no further search is possible. If the cost values are further discretized, then the tree is constructed accordingly, as described in Section~\ref{cons}. This variant of {\tt ApproxGrp} is referred to as {\tt Cons-k-Cost-ApproxGrp}.



Upon returning a non-empty subset $\mathcal{U}''$ of $\mathcal{U}'$, \\ {\tt GrpCandidateSet} terminates. Then, {\tt ApprxGrp} stores that $\alpha$ and associated $\mathcal{U}''$ and continues its binary search over $\mathcal{L}$ for a different $\alpha$. Once the binary search  ends, it returns that $\mathcal{U}''$ which has the smallest $\alpha$ associated as the solution  with the diameter upper-bounded by $2 \alpha$, as long as the distance between the workers satisfy the triangle inequality\footnote{\small Without triangle inequality assumption, no theoretical guarantee could be ensured~\cite{fd2}.}. 
In case {\tt GrpDia()} returns an empty worker set  to the main function, the binary search continues,  until there is no more option in $\mathcal{L}$.  If there is no such $\mathcal{U}''$ that is returned by {\tt GrpDia()}, then obviously the attempt to find a worker group for the task $t$ remains unsuccessful.


The pseudo-code of the algorithm {\tt ApprxGrp()} is presented in Algorithm~\ref{alg:gen}. For the given task $t$ using the example in Section~\ref{sec:ex}, $\mathcal{L}$ is ordered as follows: $0, 0.4, 0.66, 0.85, 1.0$. The binary search process in the first iteration considers $\alpha = 0.66$ and calls  {\tt GrpDia($\alpha, \{Q_i, \forall d_i\}, C)$}. In the first iteration, {\tt GrpDia()} attempts to find a {\em star graph} (referred to Figure~\ref{stargraph}) with $u_1$ as  the center of the star. This returned graph is taken as the input along with the skill threshold of $t$ inside {\tt GrpCandidateSet()}next. For our running example, subroutine ${\tt GrpDia(0.66,1.8,1.66,1.4,2.5)}$ returns $u_1, u_3, u_4, u_6$. Now notice, these $4$ workers do not satisfy the skill threshold of task $t$ (which are respectively $1.8, 1.66,1.4$ for the $3$ domains.). Therefore, {\tt GrpCandidateSet($\mathcal{U},t$)} returns false and {\tt GrpDia()} continues to check whether a star graph centered around $u_2$ satisfies the distance threshold $0.66$. Algorithm~\ref{alg:finddia} presents the pseudocode of this subroutine. When run on the example in Section~\ref{sec:ex}, {\tt ApprxGrp()}  returns workers $u_1,u_2,u_3,u_5,u_6$ as the results with objective function value upper bounded by $\leq 2 \times 0.66$.

\begin{algorithm}[t]
\caption{Approximation Algorithm {\tt ApprxGrp()}}
\label{alg:gen}
\begin{algorithmic}[1]
\begin{small}
\REQUIRE $\mathcal{U}$, human factors for $\mathcal{U}$ and task $t$  
\STATE List $\mathcal{L}$ contains all unique distance values in increasing order
\REPEAT
\STATE Perform binary search over $\mathcal{L}$ 
\STATE For a given distance $\alpha$,  $\mathcal{U}'= {\tt GrpDia(\alpha, \{Q_i, \forall d_i\}, C)}$
\IF{$\mathcal{U}' \neq \emptyset$}
\STATE Store worker group $\mathcal{U}'$ with diameter $d \leq 2 \alpha$.
\ENDIF
\UNTIL{the search is complete}
\RETURN $\mathcal{U}'$ with the smallest $d$
\end{small}
\end{algorithmic}
\end{algorithm}

\begin{algorithm}[t]
\caption{Subroutine {\tt GrpDia()}}
\label{alg:finddia}
\begin{algorithmic}[1]
\begin{small}
\REQUIRE Distance matrix of the worker set $\mathcal{U}$, distance $\alpha$, task $t$. \
\REPEAT
\STATE for each worker $u$  
\STATE form a star graph centered at $u$, such that for each edge $u, u_j$, $dist(u_,u_j) \leq \alpha$.  Let $\mathcal{U}'$ be the set of workers in the star graph.
\STATE $\mathcal{U}''$ = {\tt GrpCandidateSet($\mathcal{U}',t$)} 
\IF{$\mathcal{U}'' \neq \emptyset$}
\STATE return $\mathcal{U}''$
\ENDIF
\UNTIL{all $n$ workers have been fully exhausted}
\RETURN $\mathcal{U}''= \emptyset$
\end{small}
\end{algorithmic}
\end{algorithm}

\vspace{-0.05in}
\begin{theorem}
Algorithm {\tt ApprxGrp} has a $2$-approximation factor, as long as the distance satisfies triangle inequality.
\end{theorem}
\vspace{-0.1in}
\begin{lemma}\label{apr}
{\tt Cons-k-Cost-ApproxGrp} is polynomial.
\end{lemma}
\vspace{-0.1in}

Both these proofs are elaborated in Section~\ref{proofs} in appendix.

\subsection{Optimal Algorithm OptGrp}\label{opt}
Subroutine {\tt GrpCandidateSet()} leaves enough intuition behind to design an instance optimal algorithm that works well in practice. It calls subroutine  {\tt GrpCandidateSet()} with the actual worker set $\mathcal{U}$ and the task $t$. For {\tt OptGrp}, the tree is constructed in depth-first-fashion inside {\tt GrpCandidateSet()} and all valid solutions from the subroutine are returned to the main function. The output of {\tt OptGrp} is that candidate set of workers returned by {\tt GrpCandidateSet()} which has the smallest largest edge. When run on the example in Section~\ref{sec:ex}, this {\tt OptGrp} returns $\mathcal{G} = \{u_1,u_2,u_3,u_5,u_6\}$ with objective function value $1.0$. 

Furthermore, when workers wages are discretized into $k$ buckets, {\tt OptGrp} could be modified as described in Section~\ref{cons} and is referred to as {\tt Cons-k-Cost-OptGrp}.

\vspace{-0.1in}
\begin{theorem}
Algorithm {\tt OptGrp} returns optimal answer.
\end{theorem}

\vspace{-0.1in}
%

\begin{lemma}
{\tt Cons-k-Cost-OptGrp} is polynomial.
\end{lemma}
\vspace{-0.1in}
Both these proofs are described in Section~\ref{proofs} in appendix.

\section{Enforcing Upper Critical Mass : SPLT}\label{stage2}

\newcommand{\stageTwo}{{\tt Group-Partition}\xspace}

When {\tt Grp} results in a large unwieldy group $\mathcal{G}$ that may struggle with collaboration, it needs to be partitioned further into a set of sub-groups in the {\tt Splt} phase to satisfy the {\em upper critical mass} ($K$) constraint. At the same time, if needed, the workers across the subgroups should still be able to effectively collaborate. Precisely, these intuitions are further formalized in the {\tt Splt} phase.



\begin{definition}
{\tt Splt:} Given a group $\mathcal{G}$, decompose it into a
 disjoint set of subgroups $(G_1, G_2, \ldots, G_x)$ such that 
$\forall_i |G_i| \leq K$, $\sum_i |G_i|  = |\mathcal{G}|$ and 
the aggregated all pair inter group distance $\Sigma_{\forall_{G_i,G_j \in \mathcal{G}}}\mathit{SumInterDist}(G_i,G_j)$ is minimized.
\end{definition}


\begin{theorem}
Problem {\tt Splt}  is NP-hard.
\end{theorem}

The proof is described in Section~\ref{proofs} in appendix.

{\bf Proposed Algorithm for Splt:} Since the ILP for {\tt Splt} can be very expensive, our primary effort remains in designing an alternative strategy that is more efficient, that allows provable bounds on the result quality. We take the following overall  direction: imagine that the output of {\tt Grp} gives rise to a large group $\mathcal{G}$ with $n'$ workers, where $n' > K$. First, we determine the number of subgroups $x$ and the number of workers in each subgroup $G_i$. Then, we attempt to find optimal partitioning of the $n'$ workers across these $x$ subgroups that minimizes the objective function. We refer to this as {\tt SpltBOpt} which is the {\em optimal balanced partitioning} of  $\mathcal{G}$. For the running example in Section~\ref{sec:ex}, this would mean creating $2$ subgroups, $G_1$ and $G_2$, with $3$ workers in one and the remaining $2$ in the second subgroup using the workers $u_1,u_2,u_3,u_5,u_6$, returned by {\tt ApprxGrp}.


For the remainder of the section, we investigate how to find {\tt SpltBOpt}. There are intuitive as well as logical reasons behind taking this direction. Intuitively, lower number of subgroups gives rise to overall smaller objective function value (note that the objective function is in fact $0$ when $x=1$). More importantly, as Lemma~\ref{bp} suggests,  under certain conditions, {\tt SpltBOpt} gives rise to provable theoretical results for the {\tt Splt} problem.  Finding the approximation ratio of {\tt SpltBOpt} for arbitrary number of partitions is deferred to future work.

\begin{lemma}\label{bp}
{\tt SpltBOpt} has 2-approximation for the {\tt Splt} problem, if the distance satisfies triangle inequality, when $x=\lceil\frac{n'}{K}\rceil= 2$.
\end{lemma}

The proof is described in Section~\ref{proofs} in appendix.



%
%

Even though the number of subgroups (aka partitions) is  $\lceil\frac{n'}{K}\rceil$ with $K$ workers in all but last subgroup, finding an optimal assignment of the $n'$ workers across those subgroups that minimizes the objective function is NP-hard. The proof uses an easy reduction from \cite{guttmann2000approximation}. We start by showing how the solution to {\tt SpltBOpt} problem could be bounded by the solution of a slightly different problem variant, known as {\tt Min-Star} problem~\cite{guttmann2000approximation}. 



\begin{definition}
{\tt Min-Star} Problem: 
Given a group $\mathcal{G}$ with $n'$ workers, out of which each of $x$ workers ($u_1,u_2,\ldots, u_x$),  represents a center of a star sub-graph (each sub-graph stands for a subgroup), the objective is to partition the remaining $n'-x$ workers into one of these $x$ subgroups $G_1,G_2,\ldots,G_x$ such that  $\sum_{i=1}^{x} k_i dist(u_i, \cup_{j \neq i} G_j)$ 
$ + \sum_{i < j} k_i k_j dist(u_i, u_j)$ is minimized, where $k_i$ is the total number of workers in subgroup $G_i$.
%
\end{definition}

Intuitively, {\tt Min-Star} problem seeks to decompose the worker set into $x$ subgroups,  such that $u_i$ is the center of a star graph for subgroup $G_i$, and for a fixed set of such workers $\{ u_1, u_2, \ldots, u_x\}$,  the contribution of $u_i$ to the objective function is proportional to the sum of distances of a star subgraph rooted at $u_i$. 

\noindent {\bf Solving {\tt Min-Star}:Algorithm {\tt Min-Star-Partition}:}
The pseudocode is listed in Algorithm~\ref{alg:findPartition} and 
additional details can be found in \cite{guttmann2000approximation}. The key insight behind this algorithm is the fact that for a fixed set of workers $\{ u_1, u_2, \ldots, u_x\}$,
the second term of the objective function $\sum_{i < j} k_i k_j dist(u_i,u_j)$ is a constant.
Furthermore, this expression could only take ${n' \choose x}$ distinct values 
corresponding to all possible combination of how the workers $\{u_1, u_2, \ldots, u_x\}$ are chosen from the group $\mathcal{G}$ with $n'$ workers.
Hence for a fixed set of workers, the objective now reduces to finding an optimal subgroups $G_1, \ldots, G_x$ that minimizes the first expression. Interestingly, this expression corresponds exactly to a special case of 
the popular {\em transportation problem} \cite{grotschel1995combinatorial}
that could be solved optimally with time complexity $O(n')$ \cite{guttmann2000approximation}. We refer to~\cite{guttmann2000approximation} for further details.

Finally, the objective function of the {\tt SpltBOpt} is computed on the optimal partition of each instance of the transportation problem, and the one with the least value is returned as output. When run using $\mathcal{G} = \{u_1,u_2,u_3,u_5,u_6\}$ from {\tt ApprxGrp}, this algorithm forms  subgroups $G_1=\{u_1,u_2,u_5\}$ and $G_2=\{u_3,u_6\}$ with objective function value $3.89$.

\begin{algorithm}[t] 
\caption{Algorithm {\tt Min-Star-Partition}}
\label{alg:findPartition}
\begin{algorithmic}[1]
\begin{small}
\REQUIRE Group $\mathcal{G}$ with $n'$ workers and upper critical mass $K$
\STATE $x = \lceil \frac{n'}{K} \rceil$
\FORALL{subset $\{u_1, \ldots, u_x\} \subset \mathcal{G}$}
\STATE Find optimal subgroups $\{G_1, \ldots, G_x\}$ for $\{u_1, \ldots, u_x\}$ by formulating it as transportation problem
\STATE Evaluate objective function for $\{G_1, \ldots, G_x\}$
\ENDFOR
\RETURN subgroups $\{G_1, \ldots, G_x\}$ with least objective function
\end{small}
\end{algorithmic}
\end{algorithm}

\vspace{-0.05in}
\begin{theorem}
Algorithm for {\tt Min-Star-Partition} has a 3-approximation for {\tt SpltBOpt} problem. 
\vspace{-0.05in}
\end{theorem}

\begin{lemma} 
{\tt Min-Star-Partition} is polynomial.
\vspace{-0.05in}
\end{lemma}

Both these proofs are described in Section~\ref{proofs} in appendix.
\vspace{-0.1in}
\section{Experiments}
\label{sec:exps}

We describe our real and synthetic data experiments to evaluate our algorithms next. The real-data experiments are conducted at AMT. The synthetic-data experiments are conducted using a parametrizable crowd simulator. 
\subsection{Real Data Experiments}\label{sec:us}
Two different collaborative crowdsourcing applications are evaluated using AMT. i) Collaborative Sentence Translation (CST), ii) Collaborative Document Writing (CDW). 

{\bf Evaluation Criteria:} - The overall study is designed to evaluate: (1) Effectiveness  of the proposed optimization model, (2) Effectiveness of affinity calculation techniques, and (3) Effect of different upper critical mass values.

{\bf Workers:} A pool of $120$ workers participate in the sentence translation study, whereas, a  different pool of $135$ workers participate in the second one. Hired workers are directed to our website where the actual tasks are undertaken.

{\bf Algorithms:} We compare our proposed solution with other baselines: (1) To evaluate the first criteria, {\tt Optimal} algorithm (in Section~\ref{sec:pbm}) is compared against an alternative {\tt Aff-Unaware} Algorithm~\cite{DBLP:journals/corr/RoyLTAD14}. The latter assigns workers to the tasks considering skill and cost but ignoring affinity. (2) {\tt Optimal-Affinity-Age} and {\tt Optimal-Affinity-Region} are two  optimal algorithms that uses two different affinity calculation methods ({\tt Affinity-Age} and {\tt Affinity-Region} respectively) and are compared against each other to evaluate the second criteria. (3) {\tt CrtMass-Optimal-K} assigns workers to tasks based on the optimization objective and varies different upper critical mass values $K$, which are also compared against each other for different $K$.

{\bf Pair-wise Affinity Calculation:} Designing complex personality test~\cite{MBTI} to compute affinity is beyond the scope of this work. We instead choose some simple factors to compute affinity that have been acknowledged to be indicative factors in prior works~\cite{yantwo}. We calculate affinity in two ways - 1) {\tt Affinity-Age}: age based calculation discretizes workers in different age buckets and assigns a value of $1$ to a worker-pair, if they fall under the same bucket, $0$ otherwise. 2) {\tt Affinity-Region}: assigns a value of $1$, when two workers are from the same country and $0$ otherwise. We continue to explore more advanced affinity calculation methods in our ongoing work.

{\bf Overall user-study design:} The overall study is conducted in $3$-stages : (1) {\em Worker Profiling}: in stage-1, we hire workers and use pre-qualification tests using ``gold-data'' to learn their skills. We also learn other human factors as described next.(2) {\em Worker-to-task Assignment}: in stage-2, a subset of these hired workers are re-invited to participate, where the actual collaborative tasks are undertaken by them.(3) {\em Task Evaluation}: in stage-3, completed tasks are crowdsourced again to evaluate their quality.

%
%

\textbf{Summary of Results:} There are several key takeaways of our user study results. First and foremost, {\em effective collaboration is central to ensuring high quality results for collaborative complex tasks} as demonstrated in Figure~\ref{fig:mod} and Table~\ref{tab:userstudy} in appendix. 
Then, we evaluate $2$ different affinity computation models in Figure~\ref{fig:aff} and the results show that people from same region collaborate more effectively, as  ``correctness'' of {\tt Optimal-Affinity-Region} outperforms {\tt Optimal-Affinity-Age}. However, nothing could be said with statistical significance for the ``completeness'' dimension. Both these dimensions are suggested to be indicative in prior works~\cite{yantwo}. Interestingly, upper critical mass also has a significance in collaboration effectiveness, consequently, in the quality of the completed tasks, as shown in Figure~\ref{fig:cm}. Quality increases from $K=5$ to $K=7$, but it decreases with statistical significance when $K=10$ for {\tt CrtMass-Optimal-$10$}.   The final results of our collaborative document writing application  presented in appendix in Table~\ref{tab:userstudy} and in Section~\ref{resultadd} hold similar observations. 



\subsubsection{Stage 1 - Worker Profiling}
We hire two different sets of workers for sentence translation and document writing. The workers are informed that a subset of them will be invited (through email) to participate in the second stage of the study.

{\bf Skill learning for Sentence Translation:}  We hire $60$ workers and present each worker with a $20$ second English video clip, for which we have the ground truth translation in 4 different languages: English, French, Tamil, Bengali. We then ask them to create a translation in one of the languages (from the last three) that they are most proficient in. We measure each workers individual skill using Word Error Rate(WER)~\cite{Klakow:2002:TCW:638078.638080}.\\
{\bf Skill learning for Document Writing:}
For the second study CDW , we hire a different set of $75$ workers.
We design a  ``gold-data'' set that has $8$  multiple choice questions per task, for which the answers are known (e.g. for the MOOCs topic -  one question was, \emph{``Who founded Coursera?''}). 
The skill of each worker is then calculated as the percentage of her correct answers. For simplicity, we consider only one skill domain for both applications.\\ 
{\bf Wage Expectation of the worker:} We explicitly ask question to each worker on their expected monetary incentive, by giving them a high level description of the tasks that are conducted in the second stage of the study. Those inputs are recorded and used in the experiments.\\
{\bf Affinity of the workers:}  Hired workers are directed to our website, where they are asked to provide $4$ simple socio-demographic information: gender, age, region, and highest education. Workers anonymity is fully preserved. From there, affinity between the worker is calculated using, {\tt Affinity-Age} or {\tt Affinity-Region}.

Figure~\ref{fig:phase1} and Figure~\ref{fig:phase1_st} in appendix contain detailed workers profile distribution information.

\subsubsection{Stage 2 - Worker-to-Task Assignment}
Once the hired workers are profiled, we conduct the second and most important stage of this study, where the actual tasks are conducted collaboratively.\\ 
{\em Collaborative Sentence Translation(CST)}: We carefully choose three English documentaries of suitable complexity and length of about $1$ minute for creating subtitle in three different languages - French, Tamil, and Bengali. These videos are chosen from YouTube with titles: (1) Destroyer, (2) German Small Weapons, (3)British Aircraft TSR2. 


{\em Collborative Document Writing (CDW)}: Three different topics are chosen for this application: 1) MOOCs and its evolution, 2) Smart Phone and its evolution, 3) Top-10 places to visit in the world.  

For simplicity and ease of quantification, we consider that each task requires only one skill (ability to translate from English to one of the three other languages for CST, and expertise on that topic for CDW). The skill and cost requirements of each tasks are described in the Table~\ref{tasks}. These values are set by involving  domain experts and discussing the complexity of the tasks with them.
\begin{table}
\begin{center}
    \begin{tabular}{ | l | l | l | l|}
    \hline
    Task Name & Skill & Cost & Critical Mass  \\ \hline
    CST1- Destroyer  &  3.0 & \$5.0 & 5,7,10   \\ \hline  
    CST2- German Weapons  & 4.0 & \$5.0 & 5,7,10  \\ \hline  
    CST3 - British Aircraft & 3 & \$4.5 & 5,7,10 \\ \hline 
    CDW1- MOOCs & 5 & \$3 & 5,7,10  \\ \hline
    CDW2- Smartphone & 5 & \$3 & 5,7,10   \\ \hline
    CDW3- top-10 place & 5 & \$3 & 5,7,10  \\ \hline
    \end{tabular}
\end{center}
\vspace{-0.2in}
\caption{\small Description of different tasks; the default upper critical mass value is $5$. Default affinity calculation is region based.}\label{tasks}
\end{table}

{\bf Collaborative Task Assignment for CST:} 
We set up $2$ different worker groups per task and compare two algorithms {\tt Optimal-CST} {\tt Aff-Unaware-CST} to evaluate the effectiveness of proposed optimization model. We set up additional $2$ different worker groups for each task to compare {\tt Optimal-Affinity-Region} with {\tt Optimal-Affinity-Age}. Finally, we set up $3$ additional groups per task to compare the effectiveness of critical mass and compare {\tt CrtMass-Optimal-$5$}, {\tt CrtMass-Optimal-$7$}, {\tt CrtMass-Optimal-$10$}. This way, a total of $15$ groups are created. We instruct the workers to work incrementally using other group members contribution and also leave comment as they finish the work. These sets of tasks are kept active for $3$ days. 


{\bf Collaborative Task Assignment for CDW:} 
An similar strategy is adopted to collaboratively edit a document within $300$ words, using the quality, cost, and critical mass values of the document editing tasks, described in Table~\ref{tasks}. Workers are suggested to use the answers of the Stage-1 questionnaires as a reference.

\subsubsection{Stage 3 - Task Evaluation}
Collaborative tasks, such as knowledge synthesis, are often subjective. An appropriate technique to evaluate their quality is to leverage the \emph{wisdom of the crowds}. This way a diverse and large enough group of individuals can accurately evaluate information to nullify individual biases and the herding effect. Therefore, in this stage we \emph{crowdsource the task evaluation} for both of our applications. 

\begin{figure*}[t]
        \centering
        \begin{subfigure}[a]{0.24\textwidth}
                \includegraphics[width=\textwidth]{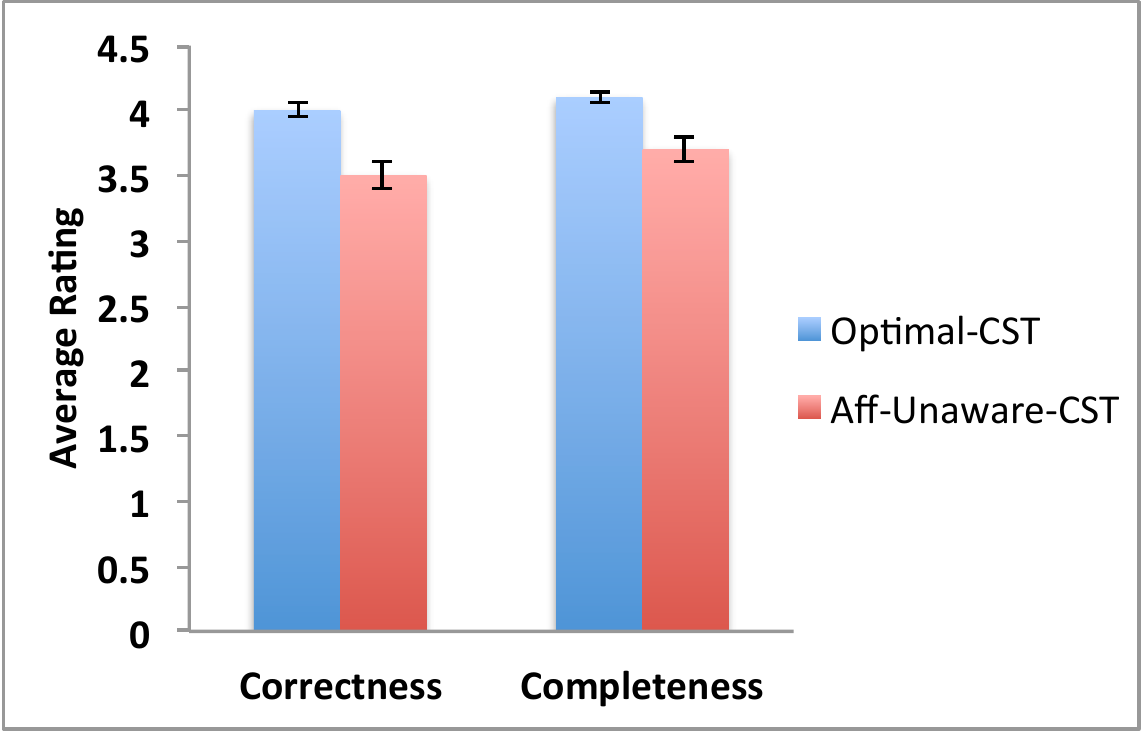}
\vspace{-0.05in}                
                \caption{Optimization Model}
                \label{fig:mod}
        \end{subfigure}
        \begin{subfigure}[a]{0.24\textwidth}
                \includegraphics[width=\textwidth]{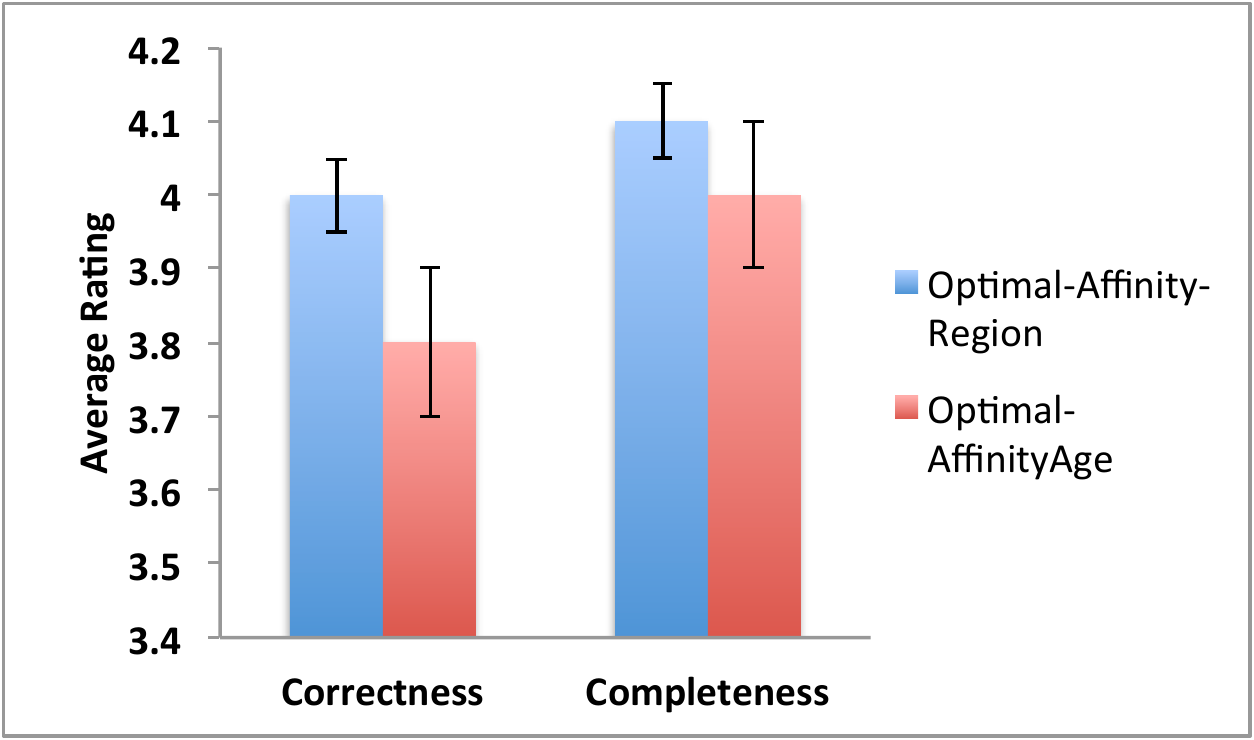}
\vspace{-0.05in}                 
                \caption{Affinity Calculation}
                \label{fig:aff}
        \end{subfigure}
        \begin{subfigure}[c]{0.24\textwidth}
                \includegraphics[width=\textwidth]{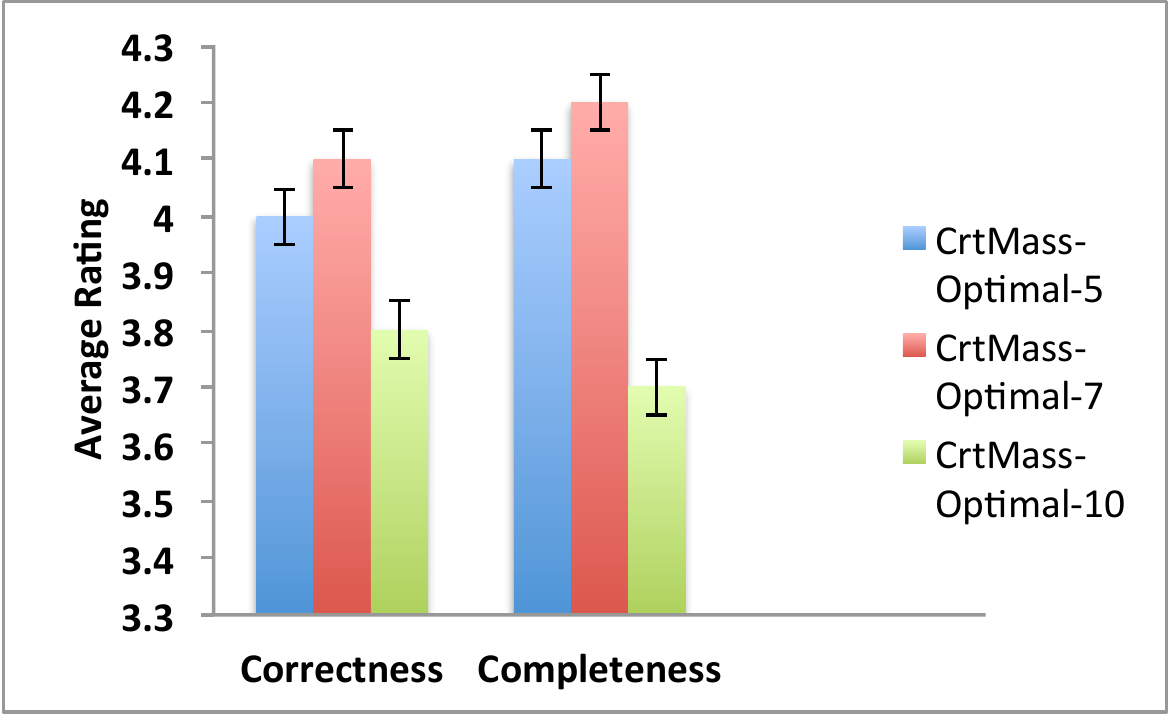}
\vspace{-0.05in}     
                \caption{Upper Critical Mass}
                \label{fig:cm}
        \end{subfigure}
        \begin{subfigure}[c]{0.25\textwidth}
                \includegraphics[width=\textwidth]{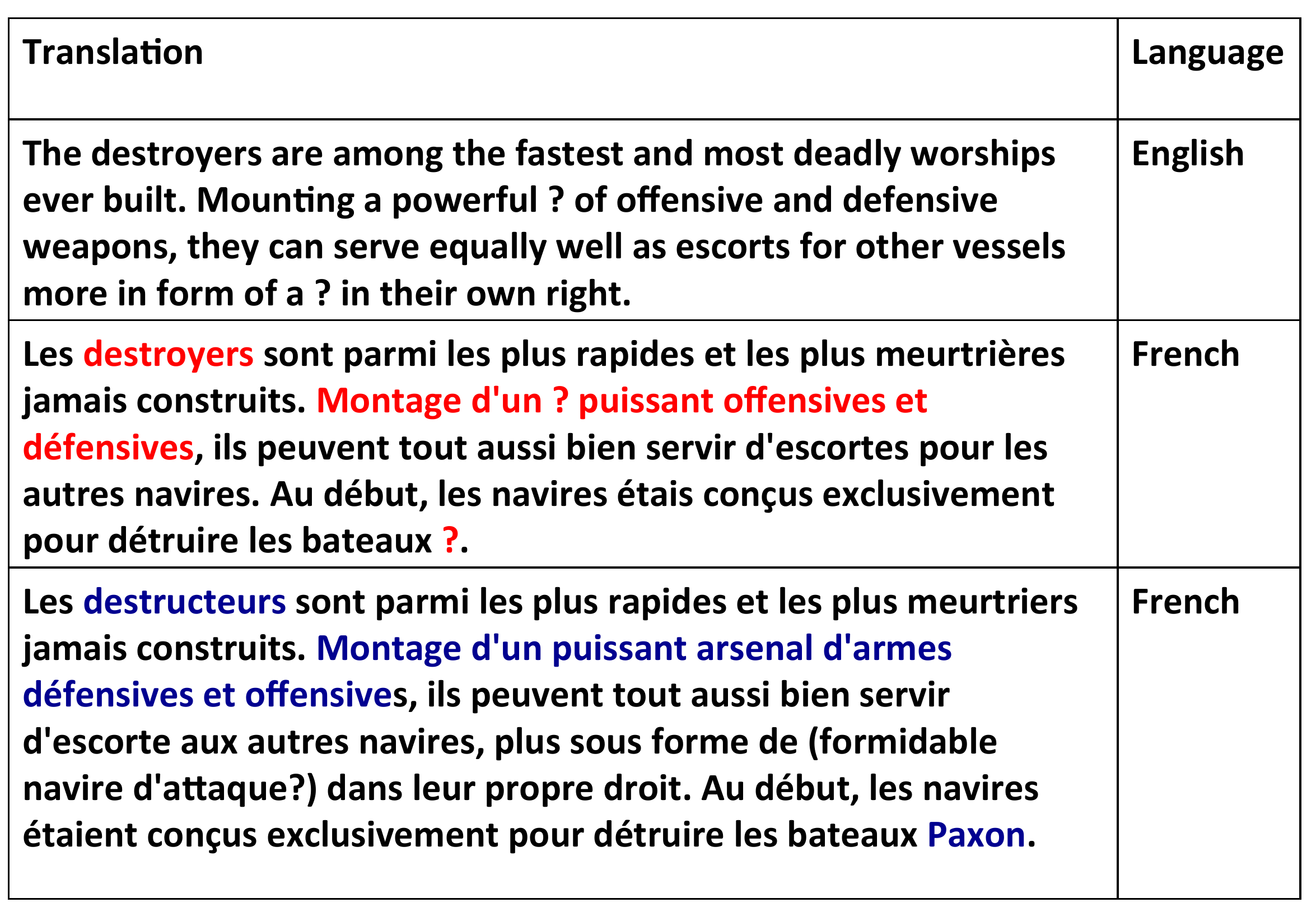}
\vspace{-0.05in}     
                \caption{A French Translation Sample}
                \label{fig:ex}
        \end{subfigure}
          \caption{\small {\bf Stage 3 results of sentence translation:} Collected data with statistical significance (standard error) is presented. These results clearly corraborate that our affinity-aware optimization model {\tt Optimal-CST} outperforms its affinity-unaware counterpart [43] with statistical significance across both quality dimensions.{\tt Optimal-Affinity-Region} apperas to outeprform {\tt Optimal-Affinity-Age} in ``correctness''. The results of {\tt CrtMass-Optimal-$10$} clearly appers to be less effective than the other two, showing some anecdotal evidence that group size is important in collaborative  crowdsourcing applications.}\label{fig:cst}
\end{figure*}

For the first study Sentence Translation (CST), we have taken $15$ final outcomes of the translation tasks as well as the original video clips and they are set up as $3$ different HITs in AMT. The first HIT is designed to evaluate the optimization model, the second one to evaluate two different affinity computation models, and the final one to evaluate the effectiveness of upper critical mass. We assign $20$ workers in each HIT, totaling $60$ new workers. Completed tasks are asked to evaluate in two quality dimensions, as identified by prior work~\cite{yantwo} - 1. correctness of translation. 2.completeness of translation. The workers are asked to rate the quality in a scale of $1-5$ (higher is better) without knowing the underlying task production algorithm. Then, we average these ratings which is similar to obtaining the viewpoint of the average readers. The CST results of different evaluation dimensions are presented in Figure~\ref{fig:cst}.

A similar strategy is undertaken for the CDW application, but the quality is assessed using $5$ key different quality aspects, as proposed in prior work~\cite{chai2009content}. For lack of space, we present a subset of these results in Section~\ref{resultadd} of the appendix in Table~\ref{tab:userstudy}. Both these results indicate that, indeed, our proposed model successfully incorporates different elements that are essential to ensure high quality in collaborative crowdsourcing tasks.

\vspace{-0.1in}
\subsection{Synthetic Data Experiments} 
We conduct our synthetic data experiments on an Intel core I5 with 6 GB RAM. We use IBM CPLEX 12.5.1 for the ILP. A  crowd simulator is implemented in Java to generate the crowdsourcing environment. All numbers are presented as the average of three runs.

{\bf Simulator Parametrization:} The simulator parameters presented below are chosen akin to their respective distributions, observed in our real AMT populations. 

\noindent 1. {\em Simulation Period} - We simulate the system for a time period of $10$ days, i.e. 14400 simulation units, with each simulation unit corresponding to 1 minutes. With one task arriving in every $10$ minutes, our default setting runs 1 day and has $144$ tasks.\\ 
\noindent 2. {\em \# of Workers} - default is $100$, but we vary $|\mathcal{U}|$ upto 5000 workers.\\
\noindent 3. {\em Workers skill and wage} - The variable $u_{d_i}$ in skill $d_i$ receives a random value from a normal distribution with the mean set to $0.8$ and a variance $0.15$. Worker's wages are also set using the same normal distribution.\\  
\noindent 4. {\em Task profile} -  The task quality $Q_{i}$, as well as cost $C$ is generated using normal distribution with specific mean $15$ and variance $1$ as default. Unless otherwise stated, each task has a skill. \\
\noindent 5. {\em Distance} - Unless otherwise stated, we consider distance to be metric and generated using Euclidean distance. 
\noindent 6. {\em Critical Mass} - the default value is 7.\\
\noindent 7. {\em Worker Arrival, Task Arrival} - By default, both workers and tasks arrive following a Poisson process, with an arrival rate of $\mu =5$/minute $1$/$10$ minute, respectively.

\begin{figure*}
\centering
\begin{minipage}[t]{0.24\textwidth}
    \includegraphics[width=\textwidth]{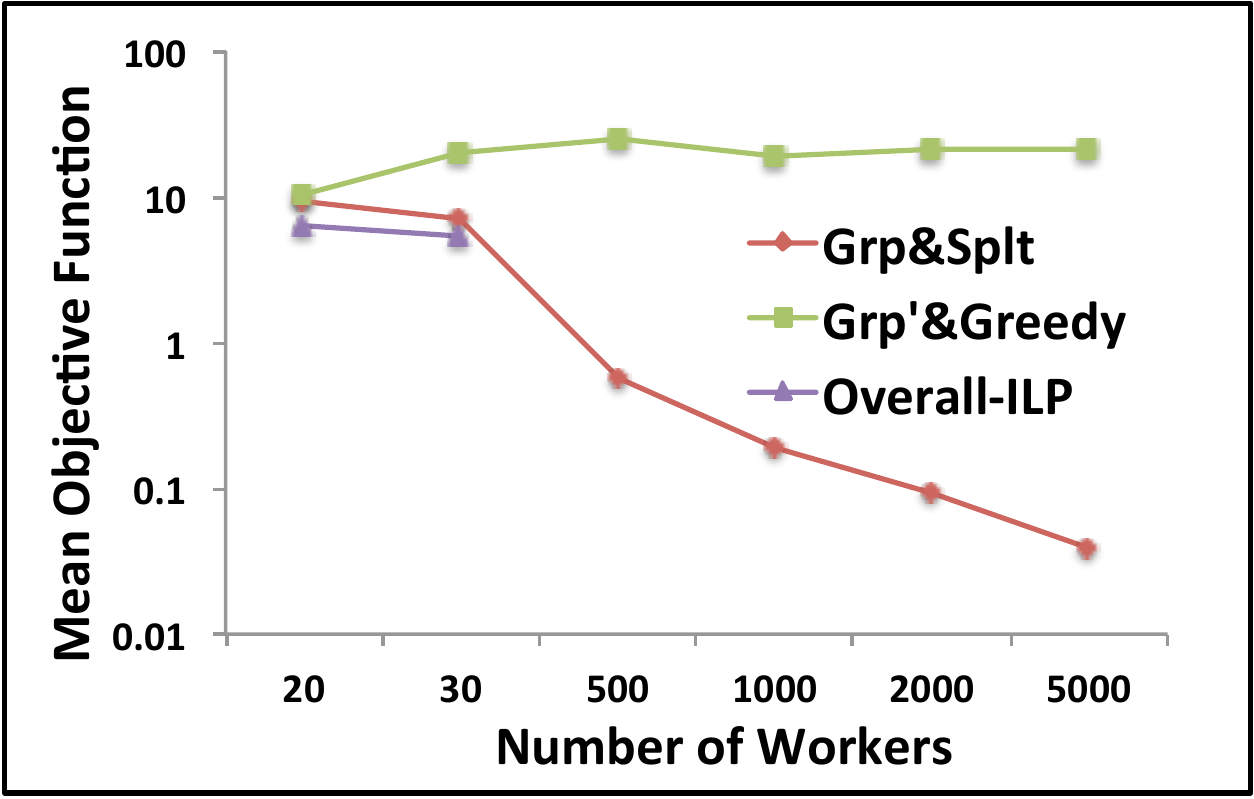}
    \caption{\small{{\tt Grp\&Splt} : Objective Function varying Number of Workers}}
    \label{fig:figObjectiveFunctionVsNumWorkers}
\end{minipage}
\begin{minipage}[t]{0.24\textwidth}
    \includegraphics[width=\textwidth]{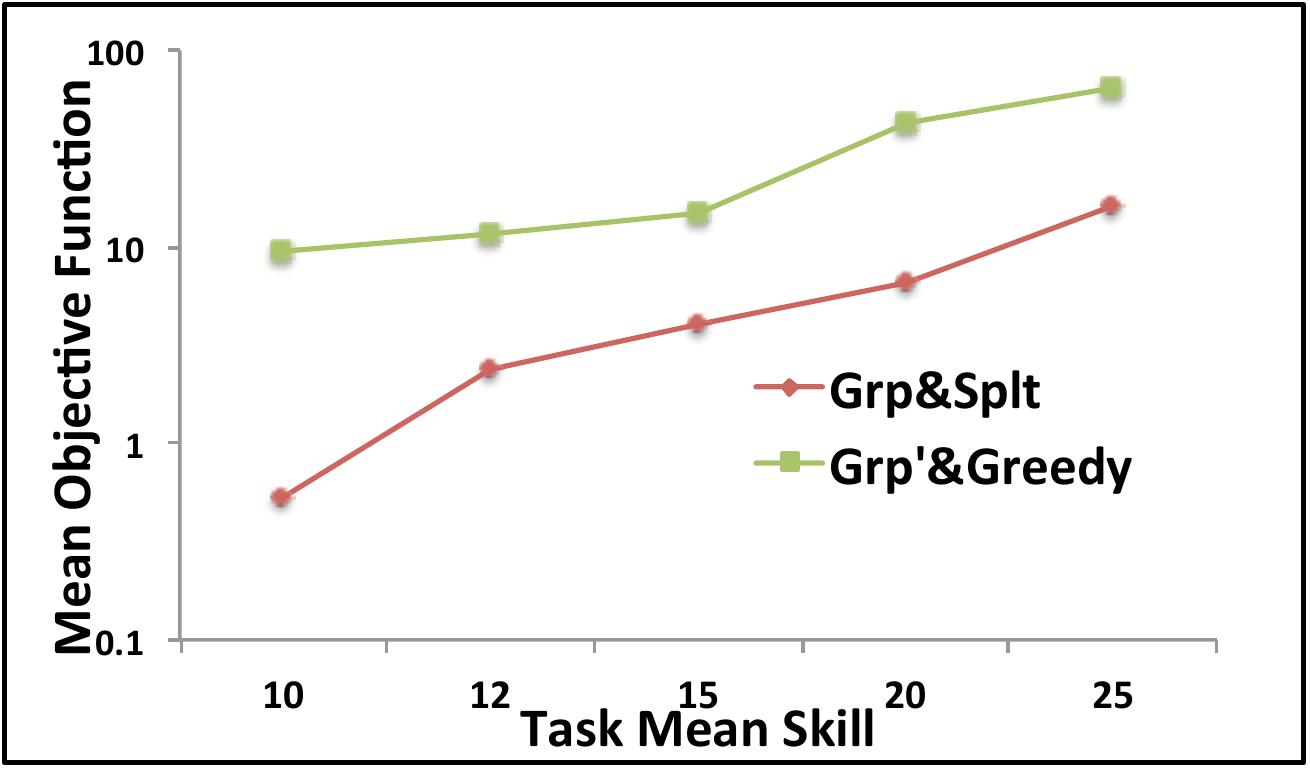}
   \caption{\small{ {\tt Grp\&Splt} : Objective Function varying Task Mean Skill}}
    \label{fig:figObjectiveFunctionVsSkill}
\end{minipage}
\begin{minipage}[t]{0.24\textwidth}
\centering
   \includegraphics[width=\textwidth]{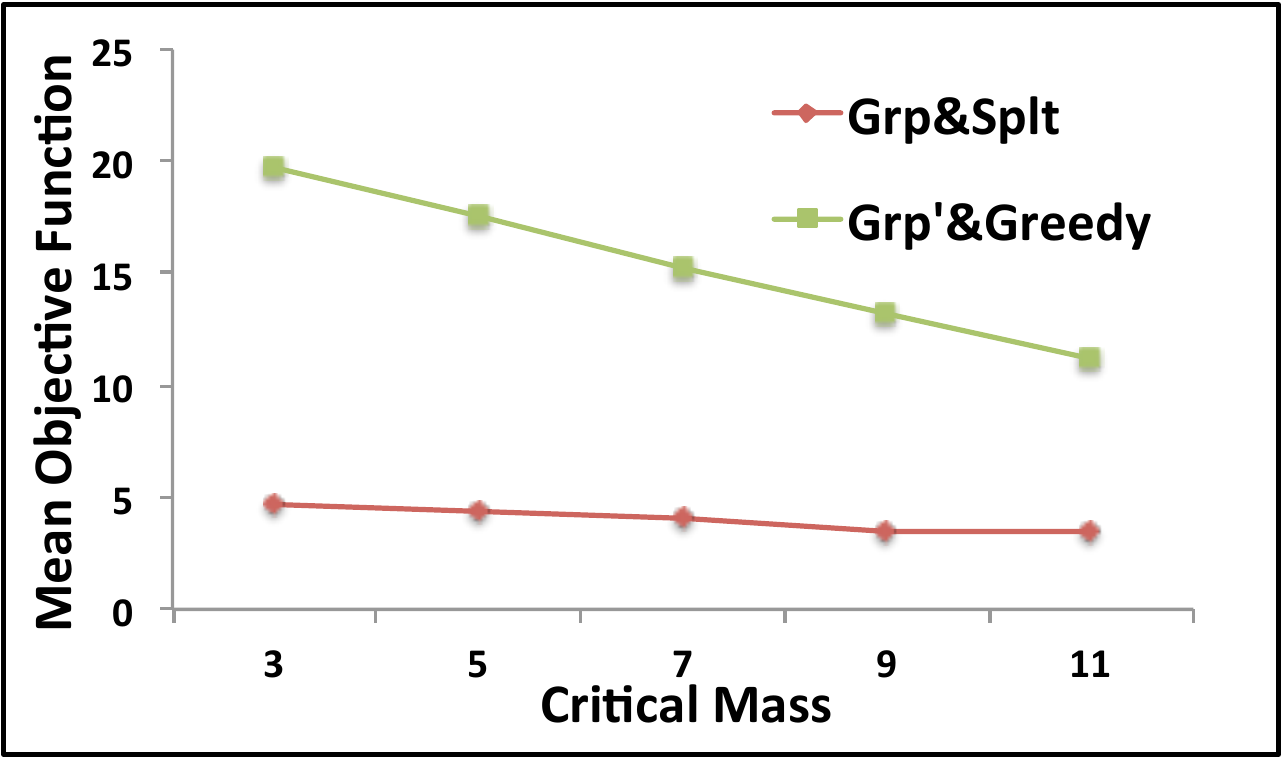}
    \caption{\small{{\tt Grp\&Splt}: Objective Function varying Critical Mass}}
    \label{fig:figObjectiveFunctionVsCriticalMass}
\end{minipage}
\begin{minipage}[t]{0.24\textwidth}
    \includegraphics[width=\textwidth]{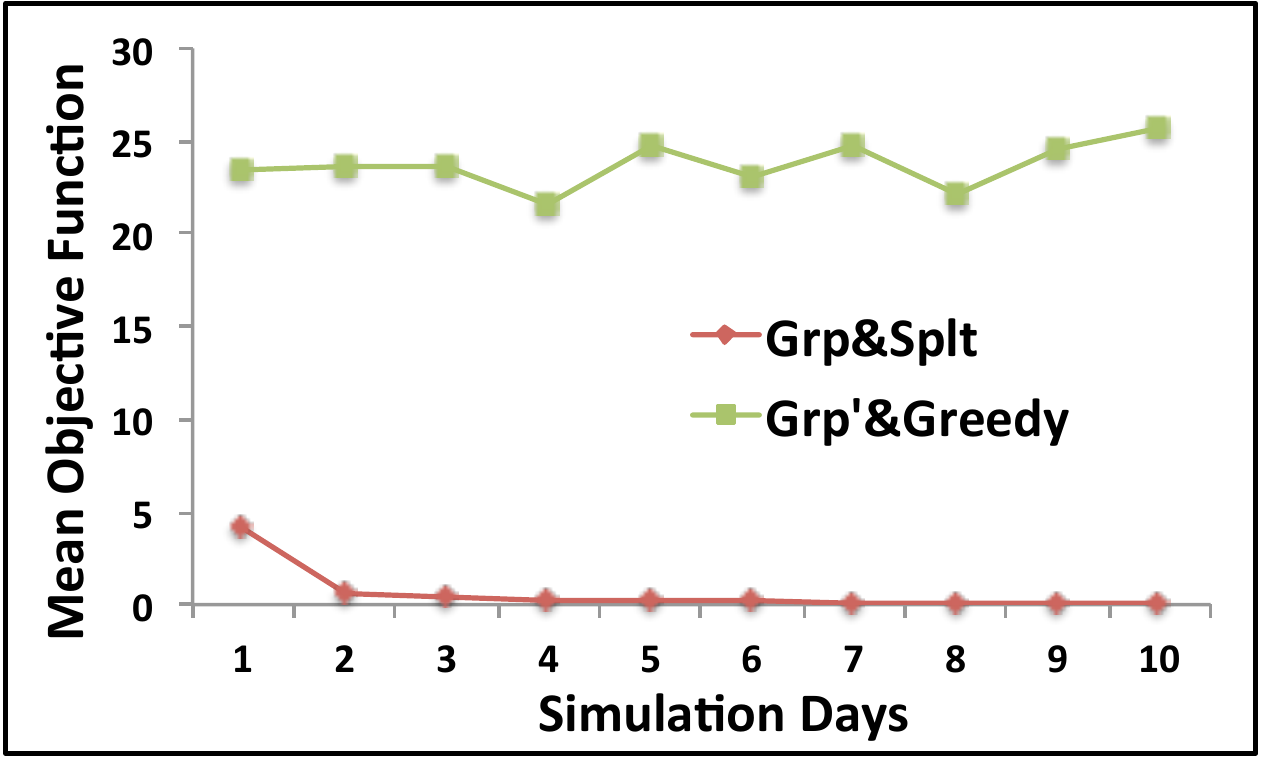}
    \caption{\small{{\tt Grp\&Splt}:Objective function over Simulation Days}}
    \label{fig:figObjectiveFunctionVsSimulationDays}
\end{minipage}
\end{figure*}

{\bf Implemented Algorithms:}
\noindent 1. {\tt Overall-ILP:} An ILP, as described in Section~\ref{sec:pbm}. \\
\noindent 2. {\tt Grp\&Splt:} Uses {\tt ApprxGrp} for {\tt Grp} and {\tt Min-Star-Partition} for {\tt Splt}.\\
\noindent 3. {\tt Grp'\&Greedy:} An alternative implementation. In phase-1, we output a random group of workers that satisfy skill and cost threshold. In phase-2, we partition users greedily into most similar subgroups satisfying critical mass constraint.\\
\noindent 4. {\tt Cons-k-Cost-ApprxGRP/Cons-k-Cost-OptGRP:} with $k=15$ as default, as discussed in Section~\ref{apx} and Section~\ref{opt}, respectively. \\
\noindent 5. {\tt GrpILP:} An ILP for {\tt Grp}.\\ 
\noindent 6. {\em No implementation of existing related work:} Due to critical mass constraint, we intend to form a group, further partitioned into a set of subgroups, whereas, no prior work has studied the problem of forming a group along with subgroups, thereby making our problem and solution unique.   

{\bf Summary of Results:} 
Our synthetic experiments also exhibit many interesting insights. First and foremost, {\tt Grp\&Splt} is a reasonable alternative formulation to solve {\tt AffAware-Crowd}, both qualitatively and efficiency-wise, as {\tt Overall-ILP} is not {\em scalable} and does not converge for more than $20$ workers. Second, our proposed approximation algorithms for {\tt Grp\&Splt} are both efficient as well as effective, and they significantly outperform other competitors. Finally, our proposed formulation {\tt AffAware-Crowd} is an effective way to optimize complex collaborative crowdsourcing tasks in a real world settings. We first present the overall quality and scalability of the combined {\tt Grp\&Splt}, followed by that of {\tt Grp} individually. Individual {\bf Splt} experiments are along the expected lines (our approach better than ILP, quality closer to optimal), and we omit those results for brevity.




\vspace{-0.05in}
\subsubsection{Quality Evaluation}
We present the quality evaluations next.
\vspace{-0.05in}
\paragraph{Grp\&Splt Quality}
The average of overall objective function value, which is the sum of $DiaDist(G)$ and aggregated all pair $SumInterDist()$ across the subgroups, is evaluated and presented as mean objective function value for $144$ tasks. {\tt Overall-ILP}  does not converge beyond $20$ workers.\\
 \noindent \textbf{\textit{Varying \# of Workers:}} Figure~\ref{fig:figObjectiveFunctionVsNumWorkers} has the results, with mean skill=$15$  and variance=$1$,  demonstrates that {\tt Grp\&Splt} outperforms {\tt Greedy-Partition} in all the cases, while being very comparable with {\tt Overall-ILP}. \\
\noindent \textbf{\textit{Varying Tasks Mean Skill:}} With varying mean skill (cost is proportional to skill), Figure~\ref{fig:figObjectiveFunctionVsSkill} demonstrates that the objective function gets higher (hence worse) for both the algorithms, as skill/cost requirement increases, while {\tt Grp\&Splt} outperforms {\tt Grp'\&Greedy}. This intuitively is meaningful, as with increasing skill requirement, the generated group is large, which decreases the workers cohesiveness further.\\
\noindent \textbf{\textit{Varying Critical Mass:}}  As Figure~\ref{fig:figObjectiveFunctionVsCriticalMass} shows, with increasing critical mass, quality of both solutions increases, because the aggregated inter-distance across the partition gets smaller due to less number of edges across. \\  
\noindent \textbf{\textit{Varying Simulation Period:}} In Figure~\ref{fig:figObjectiveFunctionVsSimulationDays} simulation period is varied, where both workers and tasks arrive based on Poisson process.  {\tt Grp\&Splt} convincingly outperforms \\ {\tt Grp'\&Greedy} in this experiment. \\
\noindent \textbf{\textit{Varying \# cost buckets:}} We also ran experiments varying the number of cost buckets for  \\ {\tt Cons-k-Cost-ApprxGRP/Cons-k-Cost-OptGRP}. With increasing $k$, the objective function gets slightly better in general.

\begin{figure*}
\centering

\begin{minipage}[t]{0.24\textwidth}
    \includegraphics[width=\textwidth]{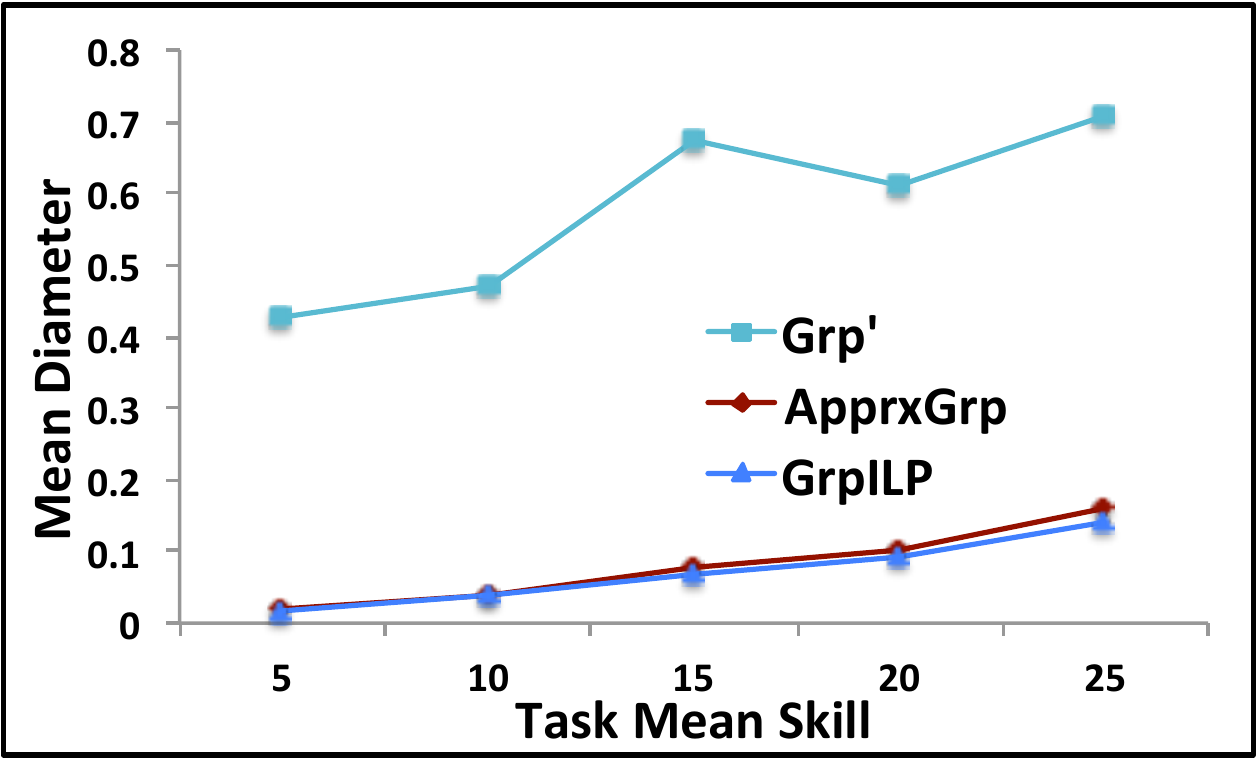}
   \caption{\small{{\tt Grp} : Mean Diameter varying Mean Skill }}
    \label{fig:figDiameterVsSkill}
\end{minipage}
\begin{minipage}[t]{0.24\textwidth}
\centering
   \includegraphics[width=\textwidth]{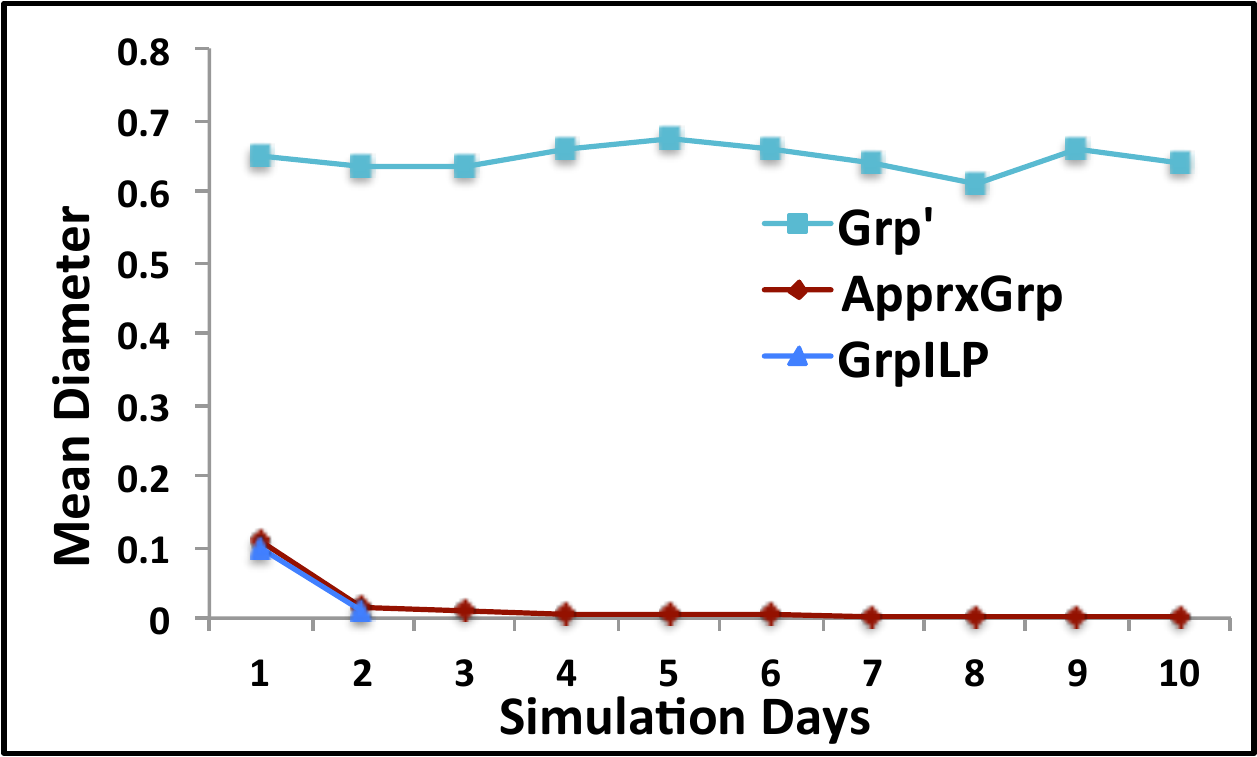}
    \caption{\small{{\tt Grp} :Mean Diamter varying Simulation Days}}
    \label{fig:figDiameterVsSimulationDays}
\end{minipage}
\begin{minipage}[t]{0.24\textwidth}
    \includegraphics[width=\textwidth]{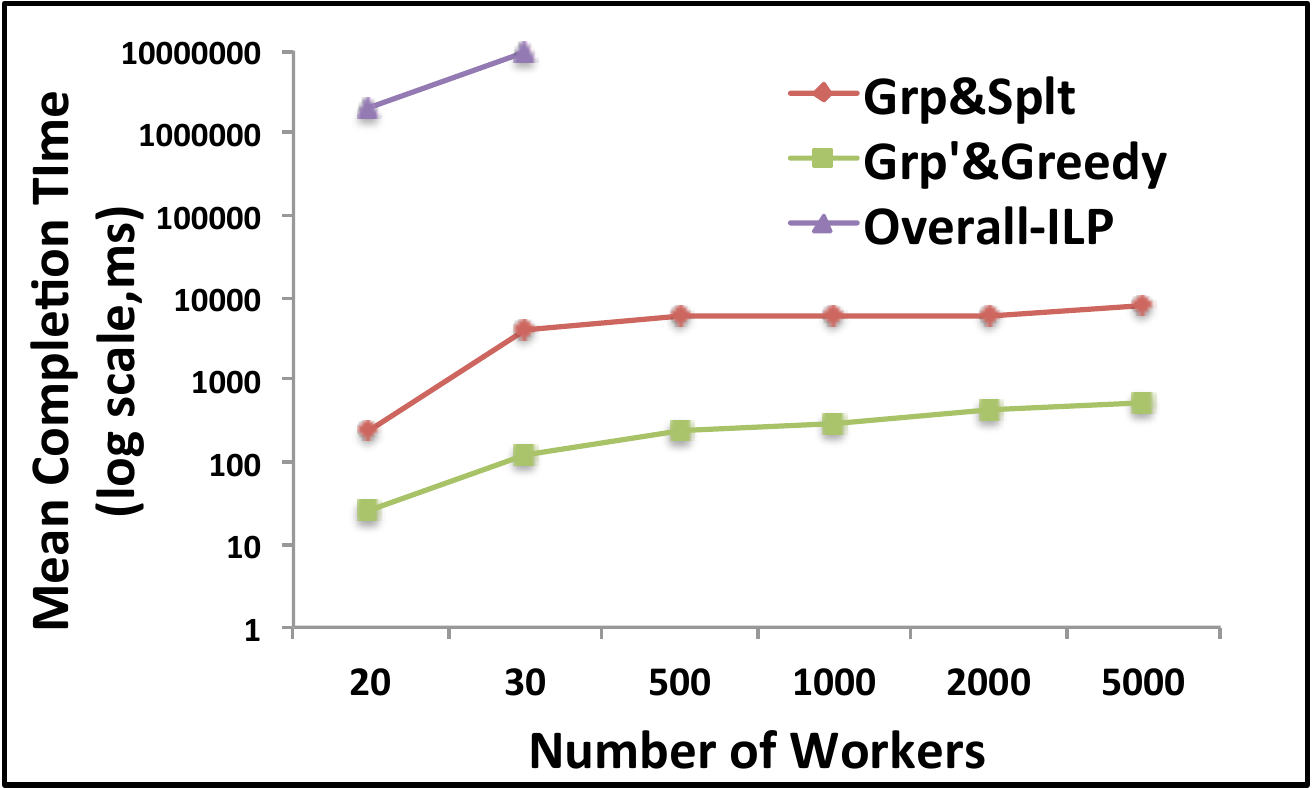}
    \caption{\small{{\tt Grp\&Splt} : Mean Completion Time varying Number of Workers}}
    \label{fig:figTimeVsNumWorkers}
\end{minipage}
\begin{minipage}[t]{0.24\textwidth}
\centering
\includegraphics[width=\textwidth]{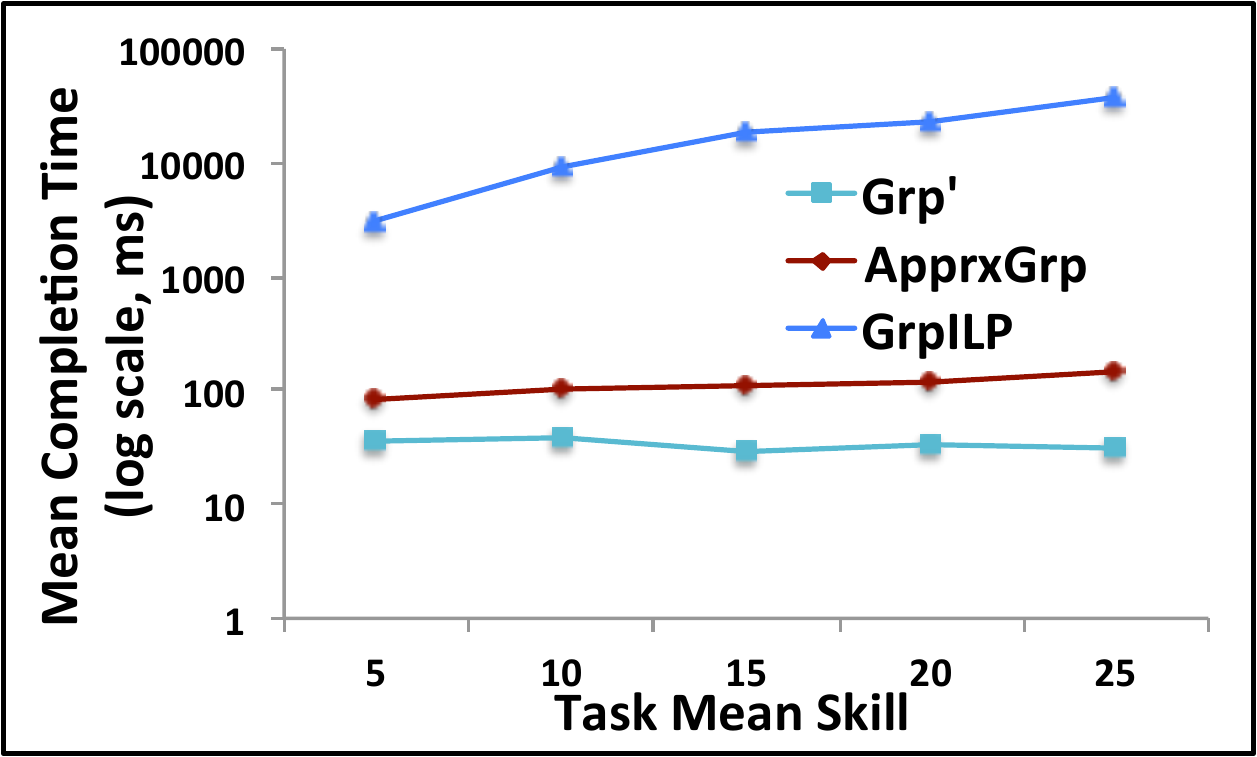}
\caption{\label{fig:figTimeVsSkill_stage1} \small{{\tt Grp} : Mean Completion Time varying Mean Skill}}
\end{minipage}
\end{figure*}

\vspace{-0.1in}
\paragraph{Grp Phase Quality}
The objective function is the average {\tt DiaDist()} value.\\
\noindent \textbf{\textit{Varying Task Mean Skill:}} Figure~\ref{fig:figDiameterVsSkill} demonstrates that, although {\tt GrpApprx} is 2-times worse than optimal theoretically,  but in practice, it is as good as optimal  {\tt GrpILP}.\\     
\noindent \textbf{\textit{Varying Simulation Period:}} Figure~\ref{fig:figDiameterVsSimulationDays} demonstrates, that, as more workers are active in the system {\tt GrpILP} cannot converge. Hence, we can not get the results for {\tt GrpILP} beyond day-2. But, {\tt GrpApprx} works fine and achieves almost optimal result.

\vspace{-0.05in}
\subsubsection{Efficiency Evaluation}
\begin{figure}
\centering
\begin{minipage}[t]{0.23\textwidth}
    \includegraphics[width=\textwidth]{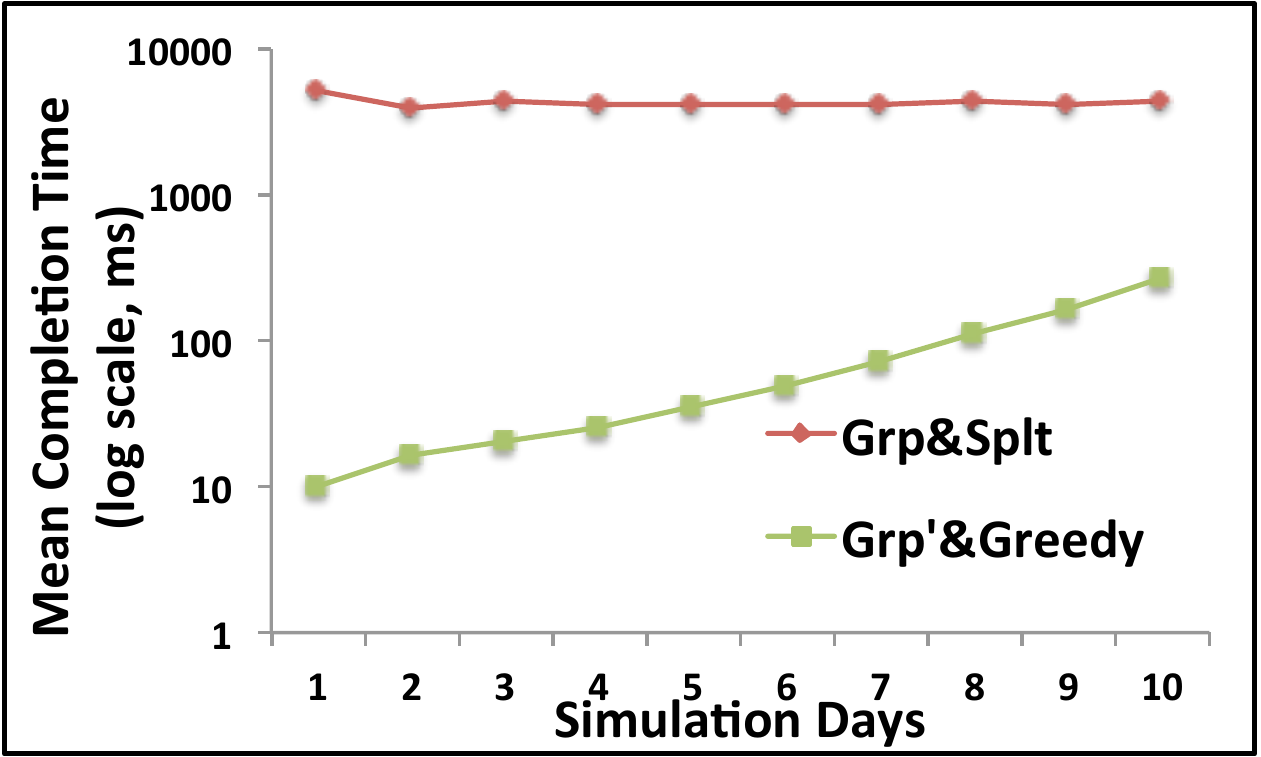}
   \caption{\small{{\tt Grp\&Splt} : Mean Completion Time varying Simulation Days}}
    \label{fig:figTimeVsSimulationDaysoverall}
\end{minipage}
\begin{minipage}[t]{0.23\textwidth}
\centering
   \includegraphics[width=\textwidth]{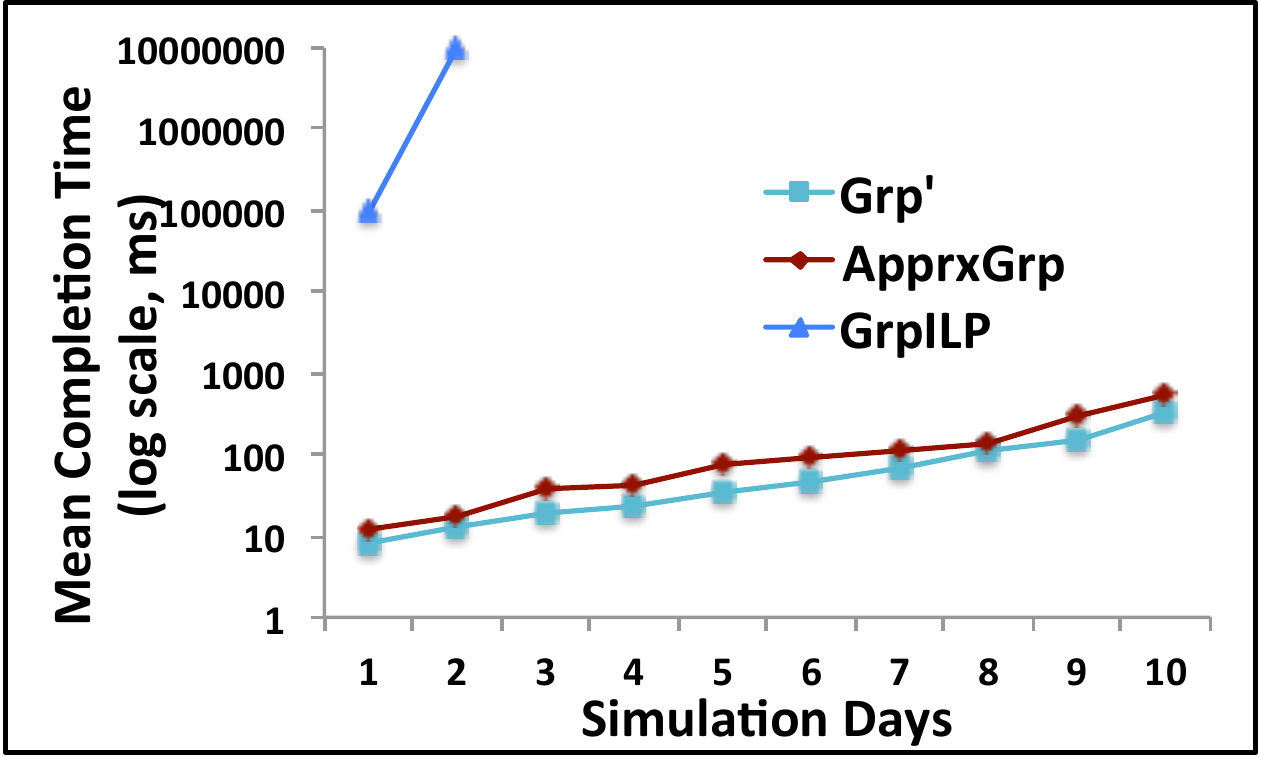}
    \caption{\small{{\tt Grp} :Mean Completion Time varying Simulation Days}}
    \label{fig:figTimeVsSimulationDays}
\end{minipage}
\end{figure}

In this section, we demonstrate the scalability aspects of our proposed algorithms and compare them with other competitive methods by measuring the average completion time of a task. Like above, we first present the overall time for both {\tt Grp} and {\tt Splt} phase, followed by only {\tt Grp} phase. 
\vspace{-0.05in}
\paragraph{Grp\&Splt Efficiency}
\noindent \textbf{\textit{Varying \# Workers:}} Figure~\ref{fig:figTimeVsNumWorkers} demonstrates that our solution {\tt Grp\&Splt} is highly scalable, whereas, {\tt Overall-ILP} fails to converge beyond 20 workers. {\tt Grp'\&Greedy} is also scalable (because of the simple algorithm in it), but clearly does not ensure high quality. \\ 
\noindent \textbf{\textit{Varying Task Mean Skill:}} Akin to previous result, {\tt Grp\&Splt} and {\tt Grp'\&Greedy} are both scalable,{\tt Grp\&Splt} achieves higher quality. We omit the chart for brevity. \\
\noindent \textbf{\textit{Varying Critical Mass:}} As before, increasing critical mass leads to better efficiency for the algorithms. We omit the chart for brevity.\\
\noindent \textbf{\textit{Varying Simulation Period:}} Figure~\ref{fig:figTimeVsSimulationDaysoverall} demonstrates that {\tt Grp\&Splt} is highly scalable in a real crowdsourcing environment, where more and more workers are entering into the system. The results show that {\tt Grp'\&Greedy} is also scalable (but significantly worse in quality). But as number of worker increases, efficiency decreases, for both, as expected.
\vspace{-0.05in}
\paragraph{Grp Phase Efficiency}
We evaluate the efficiency of  {\tt ApprxGrp} by returning mean completion time for $144$ tasks.\\
\noindent \textbf{\textit{Varying Task Mean Skill:}} As Figure~\ref{fig:figTimeVsSkill_stage1} demonstrates, {\tt ApprxGrp} outperforms {\tt GrpILP} significantly.  With higher skill threshold, the difference becomes even more noticeable. \\
\noindent \textbf{\textit{Varying Simulation Period:}}  Figure~\ref{fig:figTimeVsSimulationDays} shows the average task completion time in each day for {\tt ApprxGrp},{\tt GrpILP}, \\ {\tt Grp'\&Greedy}. Clearly, {\tt GrpILP} is impractical to use as more workers arrive in the system. 

%

\vspace{-0.1in}
\section{Related Work}
\label{sec:related}
While no prior work has investigated the problem we study here, we discuss how our work is different from a few existing works that discuss the challenges in crowdsourcing complex tasks, as well as traditional team formation problems.


{\bf Crowdsourcing Complex Tasks:} This type of human based computation \cite{Kittur:2013:FCW:2441776.2441923,Kittur:2008:HWC:1460563.1460572} handles tasks related to knowledge production, such as article writing, sentence translation, citizen science, product design, etc. These tasks are conducted in groups, are less decomposable  compared to micro-tasks (such as image tagging)~\cite{bin1,bin2}, and the quality is measured in a continuous, rather than binary scale.  

 A number of crowdsourcing tools are designed to solve application specific complex tasks. {\em Soylent} uses crowdsourcing inside a word processor to improve the quality of a written article \cite{bernstein2010soylent}. {\em Legion}, a real time user interface,  enables integration of multiple crowd workers input at the same time \cite{ Lasecki:2011:RCC:2047196.2047200}. {\em Turkit} provides an interface to programmer to use human computation inside their programming model \cite{little2010turkit} and avoids redundancy by using a {\em crash and return model} which uses earlier results from the assigned tasks. {\em Jabberwocky} is another platform which leverages social network information to assign tasks to workers and provide an easy to use interface for the programmers~\cite{ahmad2011jabberwocky}. {\em CrowdForge} divides complex task into smaller sub-tasks akin to map-reduce fashion \cite{kittur2011crowdforge}. {\em Turkomatic} introduces a framework in which workers aid requresters to break down the workflow of a complex task and thereby aiding to solve it using systematic steps \cite{kulkarni2012collaboratively}. 

Unfortunately, these related work are very targeted to specific applications and no one performs optimization based task assignment, such as ours. A preliminary work discusses modular team structures for complex crowdsourcing tasks, detailing however more on the application cases, and not on the computational challenges\cite{Retelny13}. One prior work  investigates how to assign workers to the task for knowledge intensive crowdsourcing~\cite{DBLP:journals/corr/RoyLTAD14} and its computational challenges. However, this former work does not investigate the necessity nor the benefit of {\em collaboration}. Consequently, the problem formulation and the proposed solutions are substantially different from the one studied here.

{\bf Automated Team Formation:} Although tangentially related with crowdsourcing, automated team formation is widely studied in computer assisted cooperative systems. \cite{lappas2009finding} forms a team of experts in social networks with the focus of minimizing coordination cost among team members. Although their coordination cost is akin to our affinity, but unlike us, the former does not consider multiple skills. Team formation to balance workload with  multiple skills is studied later on in~\cite{Anagnostopoulos:2010:PUF:1871437.1871515} and multi-objective optimization on coordination cost and balancing workload is also proposed  \cite{Anagnostopoulos:2012:OTF:2187836.2187950,Majumder:2012:CTF:2339530.2339690}, where coordination cost is posed as a constraint. Density based coordination is introduced in \cite{gajewar2012multi}, where multiple workers with similar skill are required in a team, such as ours. Formation of team with a leader (moderator) is studied in \cite{Kargar:2011:DTT:2063576.2063718}. Minimizing both communication cost and budget while forming a team is first considered in ~\cite{kargar2012,kargarfinding}. The concept of pareto optimal groups related to the skyline research is studied in \cite{kargar2012}.  

While several elements of our optimization model are actually adapted from these related work, there are many stark differences that precludes any easy adaptation of the team formation research to our problem. Unlike us, none of these works considers {\em upper critical mass} as a group size constraint, that forms a group multiple subgroups, which makes the former algorithms inapplicable in our settings. Additionally, none of these prior work studies our problem with the objective to maximize affinity with multiple skills and cost constraints. In \cite{chhabra2013team}, authors demonstrate empirically that the utility is decreased for larger teams which validates our approach to divide group into multiple sub-groups obeying {\em upper  critical mass}. However, no optimization is proposed to solve the problem.

In summary, principled optimization opportunities for complex collaborative tasks to maximize collaborative effectiveness under quality and budget constraints is studied for the first time in this work. 

\section{Conclusion}
\label{sec:conclusion}
We initiate the study of optimizing ``collaboration'' that naturally fits to many complex human intensive tasks. We make several contributions: we appropriately adapt various individual and group based human factors critical to the successful completion of complex collaborative tasks, and propose a  set of optimization objectives by appropriately incorporating their complex interplay. Then, we present rigorous analyses to understand the complexity of the proposed problems and  an array of efficient algorithms with provable guarantees. Finally, we conduct a detailed experimental study using two real world applications and  synthetic data to validate the effectiveness and efficiency of our proposed algorithms. Ours is one of the first formal investigations to optimize collaborative crowdsourcing. Conducting even larger scale user studies using a variety of objective functions is one of our ongoing research focus.

\appendix

\bibliographystyle{abbrv}
\bibliography{paperbib,references} 



\section{Proofs of the theorems and lemmas}\label{proofs}

{\bf Proof: Theorem 1 - {\tt AffAware-Crowd} is NP-hard.}
\begin{proof}
Sketch: 
Given a collaborative task $t$ and a set of users $\mathcal{U}$ and a real number value $X$, the decision version of the problem is, whether there is a group $\mathcal{G}$ (further partitioned into multiple subgroups) of users ($\mathcal{G} \subseteq \mathcal{U}$), such that the aggregated inter and intra distance value of $\mathcal{G}$ is $X$ and skill, cost, and critical mass constraints of $t$ are satisfied. The membership verification of the decision version of {\tt AffAware-Crowd} is clearly polynomial.

To prove NP-hardness, we consider a variant of {\em compact location}~\cite{Krumke96compactlocation} problem which is known to be NP-Complete. Given a complete graph $G$ with $N$ nodes, an integer $n \leq N$ and a real number $X'$, the decision version of the problem is whether there is a complete sub-graph $g'$ of size $n' \in N$, such that the maximum distance between between any pair of nodes in $g'$ is $X'$. This variant of the compact location problem is known as {\tt Min-DIA} in~\cite{Krumke96compactlocation}.

Our NP-hardness proof uses an instance of {\tt Min-DIA} and reduces that to an instance of {\tt AffAware-Crowd} problem in polynomial time. The reduction works as follows: each node in graph $G$ represents a worker $u$, and the distance between any two  nodes in $G$ is the distance between a pair of workers for our problem. We assume that the number of skill domain is $1$, i.e., $m=1$. Additionally, we consider that each workers $u$ has same skill value of $1$ on that domain, i.e., $u_{d} = 1,\forall u$ and their cost is $0$, i.e., $w_u = 0, \forall u$. Next, we describe the settings of the task $t$. For our problem, the task also has the quality requirement in only one domain, which is, $Q_1$. The skill, cost, and critical mass of $t$ are, $\langle Q_{1}=n', C=0, K =\infty \rangle$. This exactly creates an instance of our problem in polynomial time. Now, the objective is to form a group $\mathcal{G}$ for task $t$ such that all the constraints are satisfied and the objective function value of {\tt AffAware-Crowd} is $X'$, such that there exists a solution to the {\tt Min-DIA} problem, if and only if, a
solution to our instance of {\tt AffAware-Crowd} exists.
\end{proof}

{\bf Proof: Theorem 2 - {\tt Grp} is NP-hard.}
\begin{proof}
Sketch: Given a collaborative task $t$ with critical mass constraint and a set of users $\mathcal{U}$ and a real number $X$, the decision version of the problem is, whether there is a group $\mathcal{G}$ of users ($\mathcal{G} \subseteq \mathcal{U}$), such that the diameter is $X$, and skill and cost constraints of $t$ are satisfied.The membership verification of this decision version of  {\tt Grp} is clearly polynomial.

To prove NP-hardness, the follow the similar strategy as above. We use an instance of {\tt Min-DIA}~\cite{Krumke96compactlocation} and reduce that to an instance of {\tt Grp}, as follows: each node in graph $G$ of {\tt Min-DIA} represents a worker $u$, and the distance between any two  nodes in $G$ is the distance between a pair of workers for our problem. We assume that the number of skill domain is $1$, i.e., $m=1$. Additionally, we consider that each workers $u$ has the same skill value of $1$ on that domain, i.e., $u_{d} = 1,\forall u$ and their cost is $0$, i.e., $w_u = 0, \forall u$. Task $t$ has quality requirement on only one domain, which is, $Q_1$. The skill requirement of $t$ is $\langle Q_{1}=n'$ and cost $C=0\rangle$. 
Now, the objective is to form a group $\mathcal{G}$ for task $t$ such that the skill and cost constraints are satisfied with the diameter of {\tt Grp} as $X'$, such that there exists a solution to the {\tt Min-DIA} problem, if and only if, a
solution to our instance of {\tt Grp} exists.
\end{proof}

{\bf Proof: Theorem 3 - Algorithm {\tt ApprxGrp} has a $2$-approximation factor, as long as the distance satisfies triangle inequality.}

\begin{figure}[t]
\centering
\includegraphics[width=2.0in]{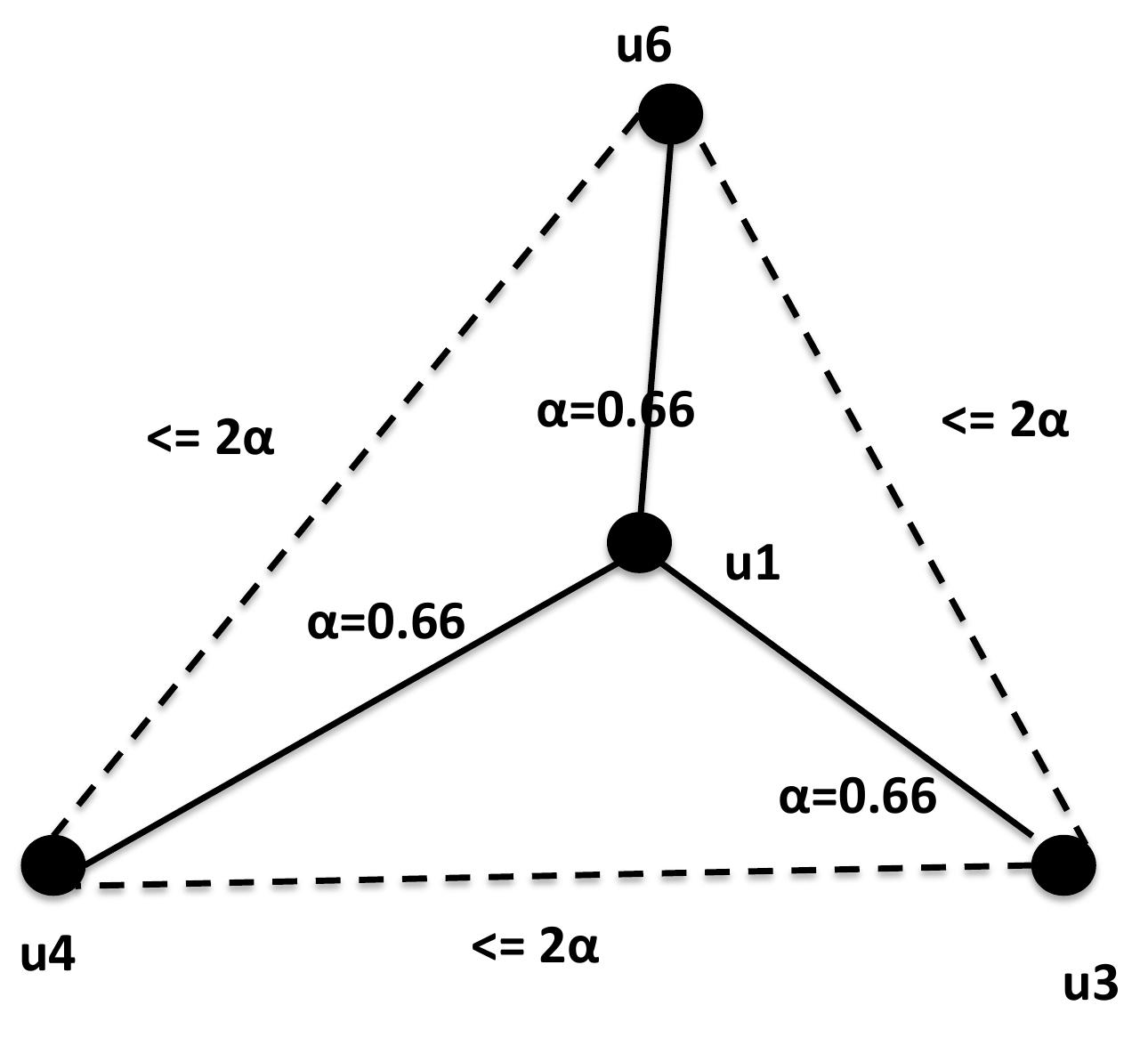}
\caption{\small \label{proof} An instantiation of ${\tt GrpDia(0.66)}$  using the example in Section~\ref{sec:ex}. The clique involving $u_1,u_3,u_4,u_6$ can not have an edge with distance $>2\times 0.66$, due to the triangle inequality property.}
\end{figure}

\begin{proof}
Algorithm {\tt ApprxGrp} overall works as follows: it sorts the distance values in ascending fashion to create a list $\mathcal{L}$ and performs a binary search over it. For a given distance value $\alpha$, it makes a call to {\tt GrpDia($\alpha$)}. Recall Figure~\ref{stargraph} that forms a star graph centered on $u_1$ with ${\tt GrpDia(0.66)}$  using the example in Section~\ref{sec:ex}. Consider Figure~\ref{proof} and notice that for a given distance value =$\alpha$, the complete graph induced by the star graph can not have any edge that is larger than $2 \times \alpha$, as long as the distance satisfies the triangle inequality property. Therefore, when {\tt GrpDia($\alpha$)} returns a non-empty worker set (that only happens when the skill and cost thresholds are satisfied), then, those workers satisfies the skill and cost threshold with the optimization objective value of $\leq 2\alpha$. Next, notice that algorithm {\tt ApprxGrp} overall attempts to return the smallest distance $\alpha$' for which {\tt GrpDia($\alpha$')} returns a non-empty set, as it performs a binary search over the sorted list of distance values (where distance is sorted in smallest to largest). Therefore, any group of workers returned by {\tt ApprxGrp} satisfies the skill and cost threshold value and $\mathit{DiaDist}(\mathcal{G})$ is at most $2$-times worse than the optimal. Hence the approximation factor holds.
\end{proof}

{\bf Proof : Lemma~\ref{apr} - {\tt Cons-k-Cost-ApproxGrp} is polynomial.}

\begin{proof}
Under a constant number of $k$-costs, subroutine {\tt GrpCandidateSet()} will accept a polynomial computation time of $O(p+1)^{mk}$ at the worst case, where $p$ is the maximum number of workers in one of the $k$ buckets ($p = O(n)$). Subroutine {\tt GrpDia()} runs for all $n$ workers at the worst case, and there is a maximum number of $log_2|\mathcal{L}|$ calls to {\tt GrpDia()} from the main function ($ |\mathcal{L}| = O(n^2)$). Therefore, the asymptotic complexity of {\tt Cons-k-ApproxGrp}  is $O(n \times log_2|\mathcal{L}| \times (p+1)^{mk})$, which is polynomial.
\end{proof}

{\bf Proof: Theorem 4 - Algorithm {\tt OptGrp} returns optimal answer.}

\begin{proof}
sketch: Algorithm {\tt OptGrp} invokes the subroutine {\tt GrpCandidateSet()}. Notice that {\tt GrpCandidateSet()} operates in the spirit of the branch-and-bound technique~\cite{lawler1966branch} to efficiently explore the search space and avoid unnecessary  computations. {\tt GrpCandidateSet()} exploits the upper bound of cost and lower bound of skill to prune out all unnecessary branches of the search tree, as shown in Figure~\ref{tree} and Figure~\ref{tree2}. However, this subroutine returns all valid worker groups to {\tt OptGrp}, and then, the main function selects the group with the smallest longest edge (i.e., smallest diameter value), and minimizes the objective function. Therefore, {\tt OptGrp} is instance optimal, i.e., it returns the group of workers with the smallest diameter distance, while satisfying the skill and cost threshold. Therefore, {\tt OptGrp} returns optimal answer.
\end{proof}
%

{\bf Proof : Lemma 2 - {\tt Cons-k-Cost-OptGrp} is polynomial.}

\begin{proof}
Under a constant number of $k$-costs, subroutine {\tt GrpCandidateSet()} will accept a polynomial computation time of $O(n+1)^{mk}$ at the worst case. Once the subroutine returns all valid answers, the main function will select the one that has the smallest diameter. Therefore, the computation time of {\tt Cons-k-Cost-OptGrp} is dominated by the computation time of the subroutine {\tt GrpCandidateSet()}. Therefore, Algorithm {\tt Cons-k-OptGrp} runs in polynomial time of $O((p+1)^{mk}$.
\end{proof}

{\bf Proof: Theorem 5 - Problem {\tt Splt}  is NP-hard.}

\begin{proof}
Given a group $\mathcal{G}$, an upper critical mass constraint $K$, and a real number $X$, the decision version of the {\tt Splt} is whether $\mathcal{G}$ can be decomposed to a set of subgroups such that the aggregated distances across the subgroups is $X$ and the size of each subgroup is $\leq K$. The membership verification of {\tt Splt} is clearly polynomial.

To prove NP-hardness, we reduce the {\tt Minimum Bisection}~\cite{mb} which is known to be NP-hard to an instance of {\tt Splt} problem. 

Given a graph $G(V,E)$ with non-negative edge weights the goal of Minimum Bisection problem is to create $2$ non-overlapping partitions of equal size, such that the total weight of cut is minimized. The hardness of the problem remains, even when the graph is complete~\cite{mb}.


Given a complete graph with $n'$ nodes, the decision version of the {\tt Minimum Bisection} problem is to see whether there exists a $2$ partitions of equal size, such that the total weight of the cut is $X'$.  We reduce an instance of {\tt Minimum Bisection} to an instance of {\tt Splt} as follows: the complete graph represents the set of workers with non-negative edges as their distance and we wish to decompose this group to two sub-groups, where the upper critical mass is set to be $K=n'/2$.  Now, the objective is to form the sub-groups with the aggregated inter-distance of $X'$, such that there exists a solution to the {\tt Minimum Bisection} problem, if and only if, a solution to our instance of {\tt Splt} exists.
\end{proof}

{\bf Proof: Lemma 3 - {\tt SpltBOpt} has 2-approximation for the {\tt Splt} problem, if the distance satisfies triangle inequality, when $x=\lceil\frac{n'}{K}\rceil= 2$.}

\begin{figure}[t]
\centering
\includegraphics[width=2.5in]{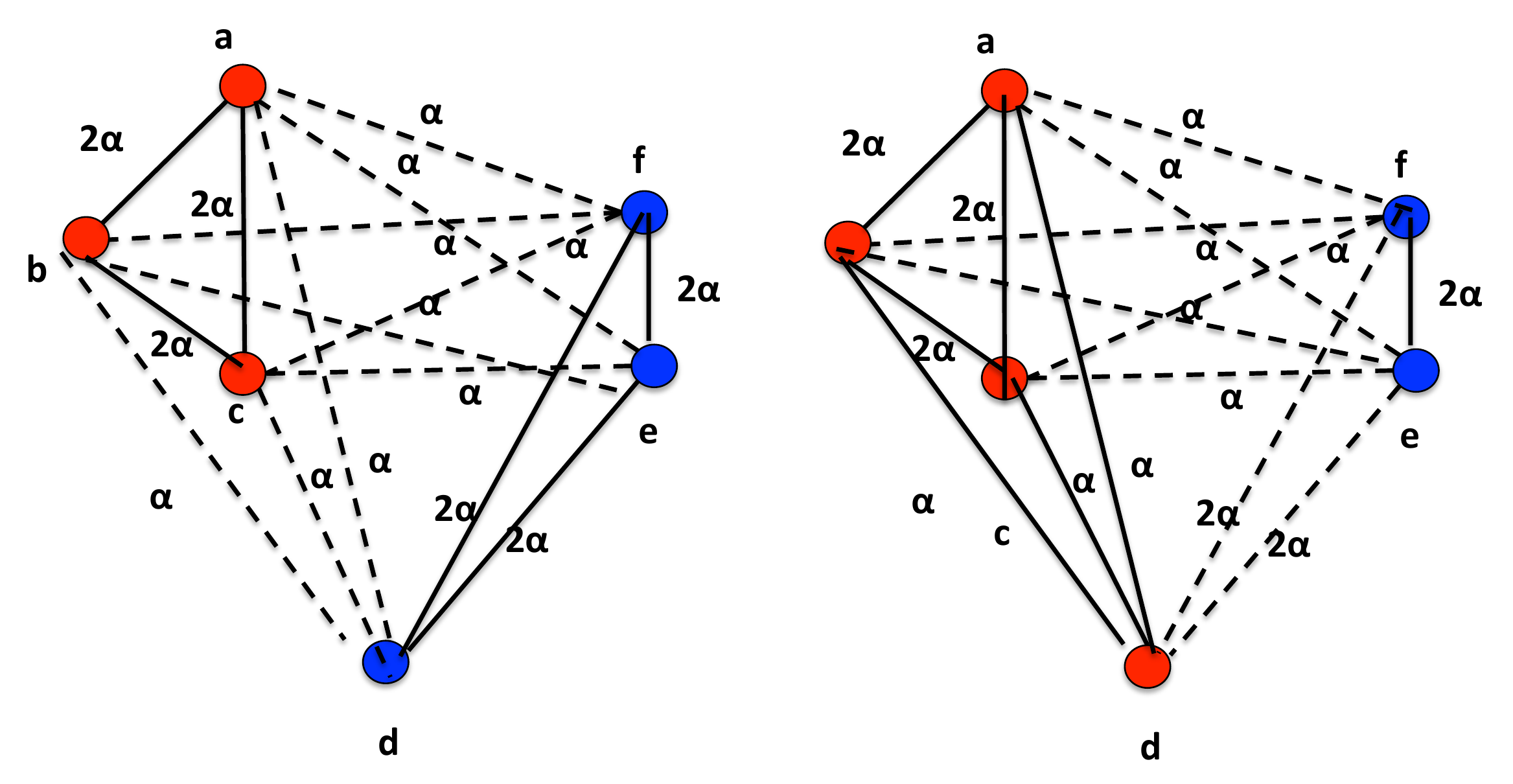}
\caption{\small \label{proof2} Balanced Partitioning in {\tt SpltBOpt} when the distance satisfies triangle inequality for a graph with $6$ modes. The left hand side figure has two partitions($\{a,b,c\}, \{d,e,f\}$) with 3-nodes in each (red nodes create one partition and blue nodes create another). The intra-partion edges are drawn solid, whereas, inter-partition edges are drawn as dashed. Assuming $K=4$, in the right hand side figure, node $d$ is moved with $a,b,c$. This increases the overall inter-partition weights, but is bounded by a factor of $2$.}
\end{figure}

\begin{proof}
Sketch: For the purpose of illustration, imagine that a graph with $n'$ nodes is decomposed into two partitions. Without loss of generality, imagine partition-1 has $n_1$ nodes and partition-2 has $n_2$ nodes, where $n_1+n_2=n'$ with total weight of $w'$. Let $K$ be the upper critical mass and assume that $K>n_1, K>n_2$. For such a scenario,  {\tt SpltBOpt} will move one or more nodes from the lighter partition to the heavier one, until the latter  has exactly $K$ nodes (if both partitions have same number of nodes then it will choose the one which gives rise to overall lower weight). Notice, the worst case happens, when some of the intra-edges with higher weights now become inter edges due to this balancing act. Of course, some inter-edges  also gets knocked off and becomes intra-edges. It is easy to notice that the number of inter-edges that gets knocked off is always larger than that of the number of  inter-edges added (because the move is always from the lighter partition to the heaver one). The next argument we make relies heavily on the triangle inequality property. At the worst case, every edge that gets added due to balancing, could at most be twice the weight of an edge that gets knocked off. Therefore, an optimal solution of {\tt SpltBOpt} has 2-approximation factor for the {\tt Splt} problem.

An example scenario of such a balancing has been illustrated in Figure~\ref{proof2}, where $n_1=n_2=3,K=4$. Notice that after this balancing, three inter-edges get deleted (ad,bd,cd), each of weight $\alpha$ and two inter-edges get added, where each edge is of weight $2 \alpha$. However, the approximation factor of $2$ holds, due to the triangle inequality property. 
\end{proof}

{\bf Proof: Theorem 6 - Algorithm for {\tt Min-Star-Partition} has a 3-approximation for {\tt SpltBOpt} problem. }

\begin{proof}
sketch: This result is a direct derivation of the previous work~\cite{guttmann2000approximation}. Previous work~\cite{guttmann2000approximation} shows that {\tt Min-Star-Partition} obtains a 3-approximation factor for the {\tt Minimum k-cut} problem. Recall that {\tt SpltBOpt} is derived from {\tt Minimum k-cut} by setting each partition size (possibly except the last one) to be equal with $K$ nodes, giving rise to a total number of $\lceil\frac{n'}{K}\rceil$ partitions. After that, the result from ~\cite{guttmann2000approximation} directly holds.
\end{proof}

{\bf Proof: Lemma 4 - {\tt Min-Star-Partition} is polynomial.}

\begin{proof}
It can be shown that {\tt Min-Star-Partition} takes $O(n'^{x+1})$ time, as there are $O(n'^x)$ distinct transportation problem instances (corresponding to each one of ${n' \choose x}$ combinations), and each instance can be solved in $O(n')$~\cite{guttmann2000approximation} time.
Since, $x$ is a constant, therefore, the overall running time is polynomial.
\end{proof}



\section{User Study Details}\label{resultadd}
This section in the appendix is dedicated to provide additional results of the user studies in Section~\ref{sec:us}. We present the partial results of distribution of workers' profile for both applications. Additionally, the Stage-2 results of collaborative document writing application is presented here.

\begin{figure*}[b]
        \centering
        \begin{subfigure}[b]{0.24\textwidth}
                \includegraphics[width=\textwidth]{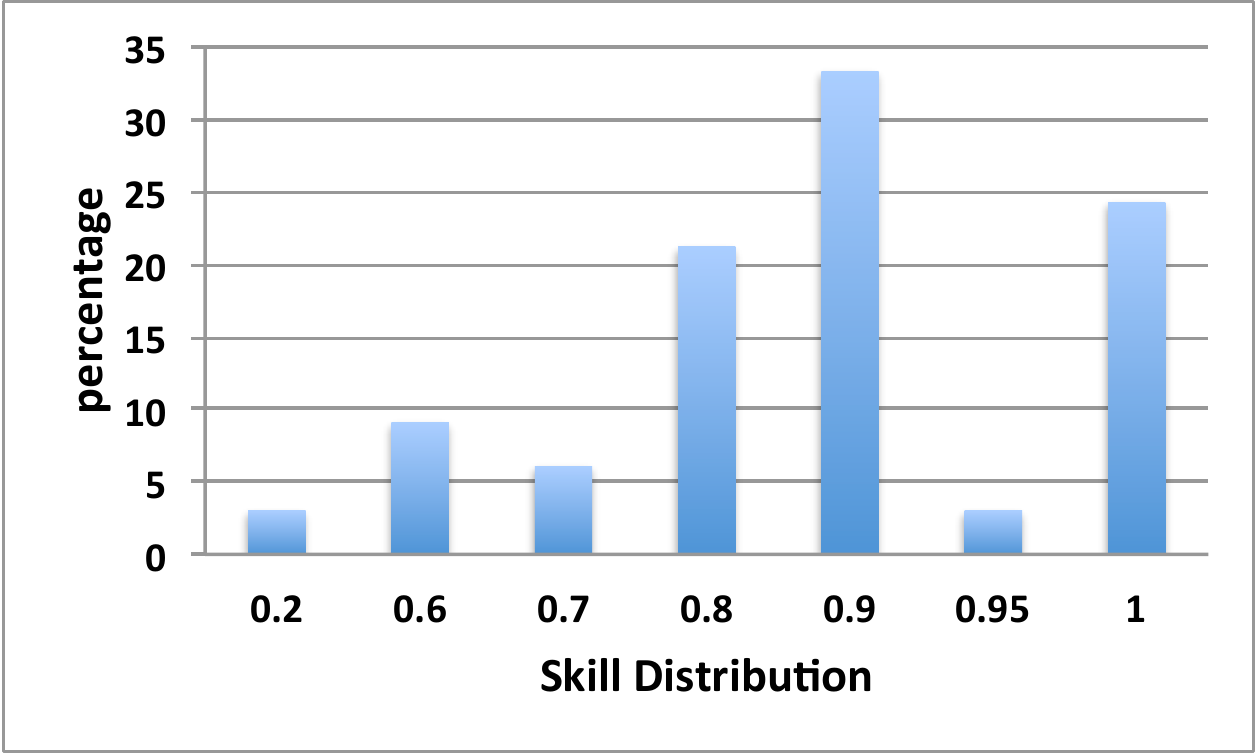}
                \caption{Worker Skill distribution}
                \label{fig:stSkill}
        \end{subfigure}
        \begin{subfigure}[b]{0.24\textwidth}
                \includegraphics[width=\textwidth]{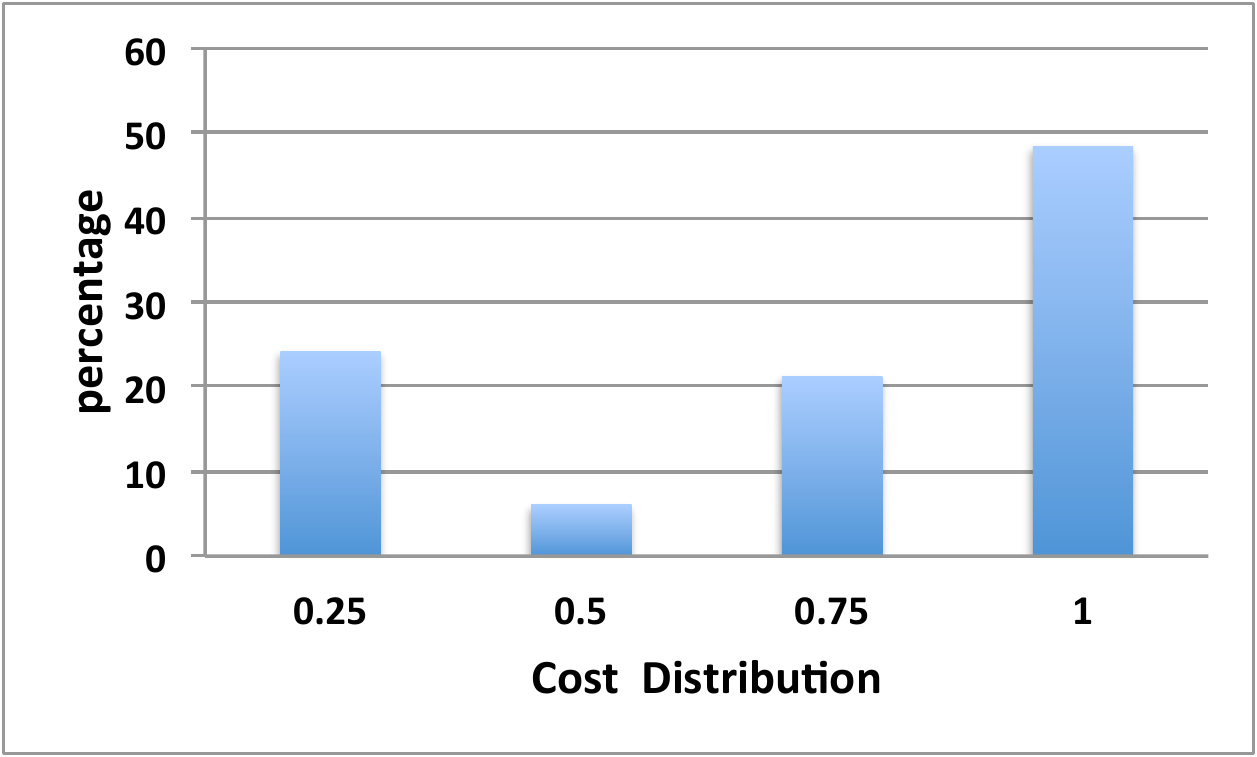}
                \caption{Worker wage distribution}
                \label{fig:stWage}
        \end{subfigure}
        \begin{subfigure}[b]{0.24\textwidth}
                \includegraphics[width=\textwidth]{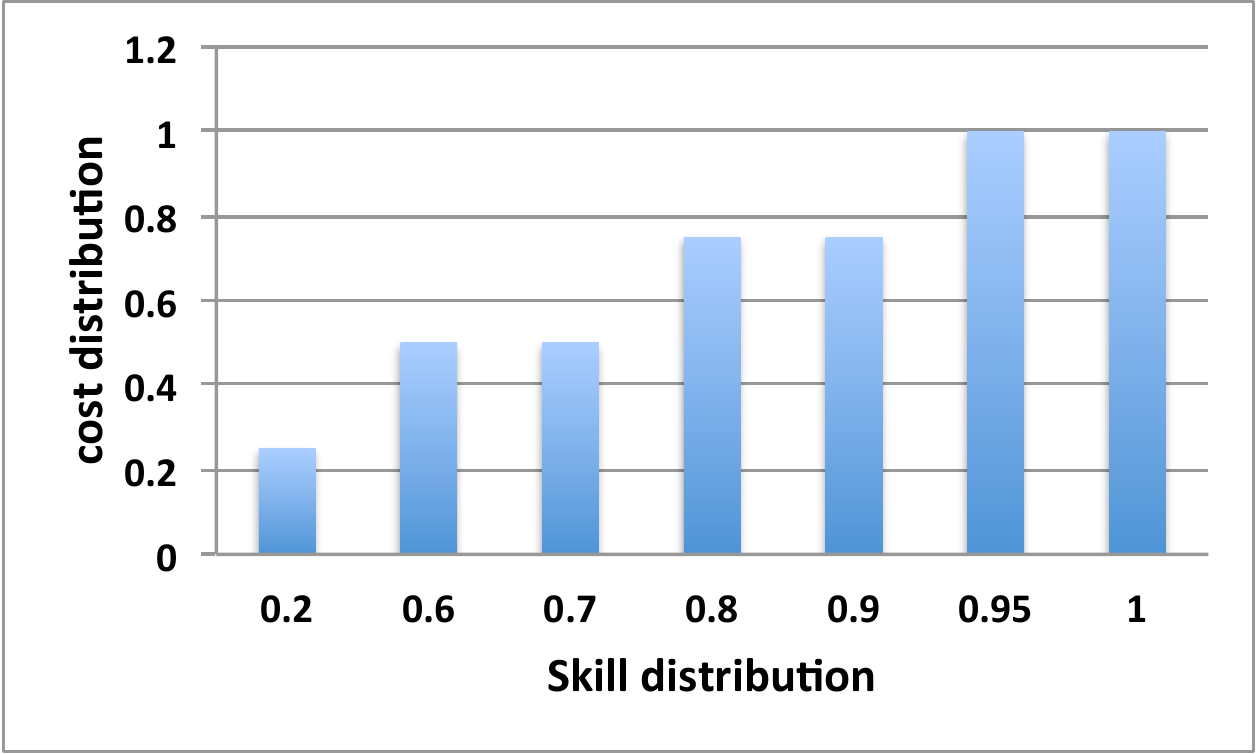}
                \caption{Distance Distribution-Region}
                \label{fig:stRegion}
        \end{subfigure}
         \begin{subfigure}[b]{0.24\textwidth}
                \includegraphics[width=\textwidth]{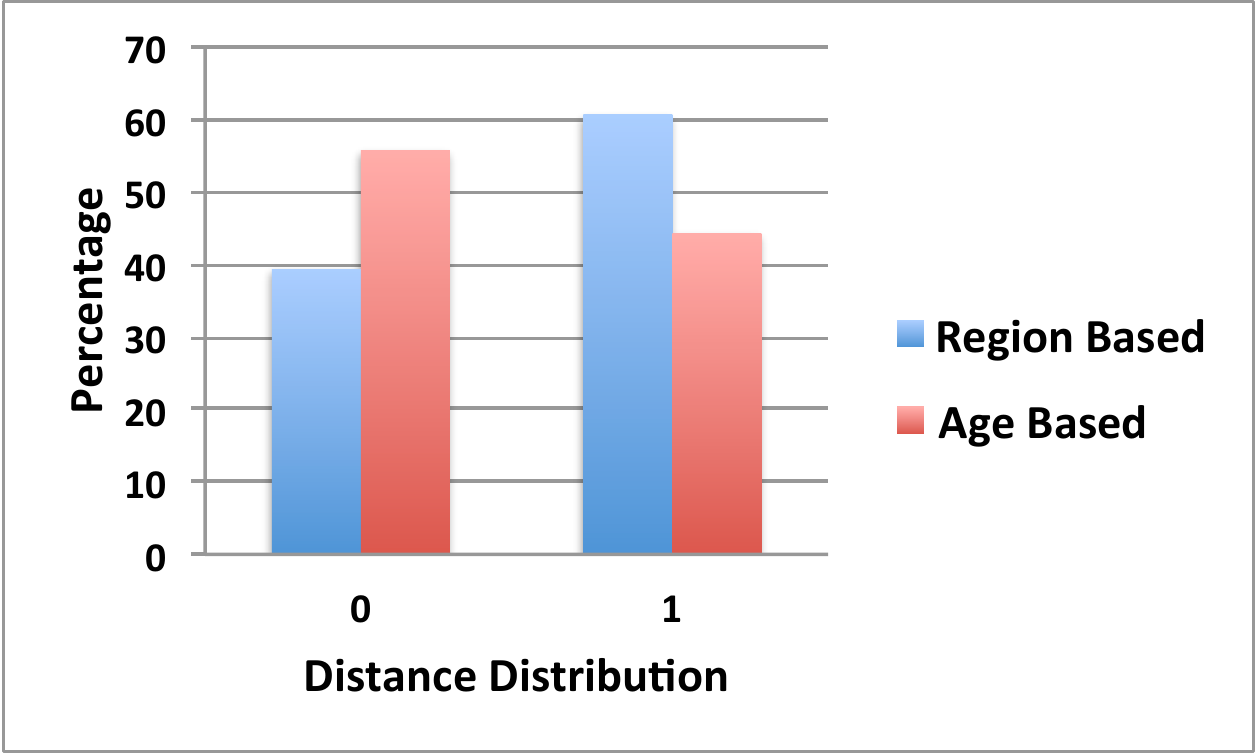}
                \caption{Distance Distribution-Age}
                \label{fig:stAge}
        \end{subfigure}
       \caption{Worker profile distributions for the Sentence Translation Tasks in Section~\ref{sec:us}}\label{fig:phase1_st}
\end{figure*}

\begin{figure*}[b]
        \centering
        \begin{subfigure}[b]{0.24\textwidth}
                \includegraphics[width=\textwidth]{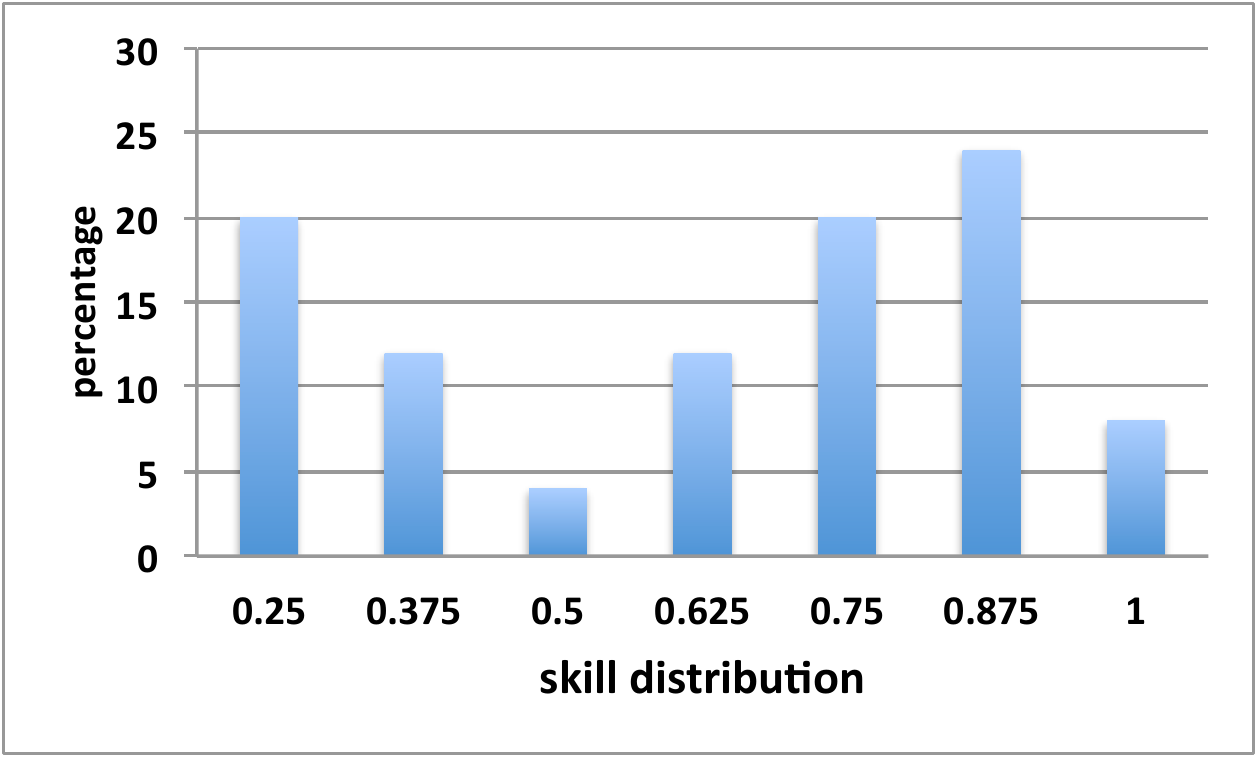}
                \caption{Worker Skill distribution}
                \label{fig:scn1}
        \end{subfigure}
        \begin{subfigure}[b]{0.24\textwidth}
                \includegraphics[width=\textwidth]{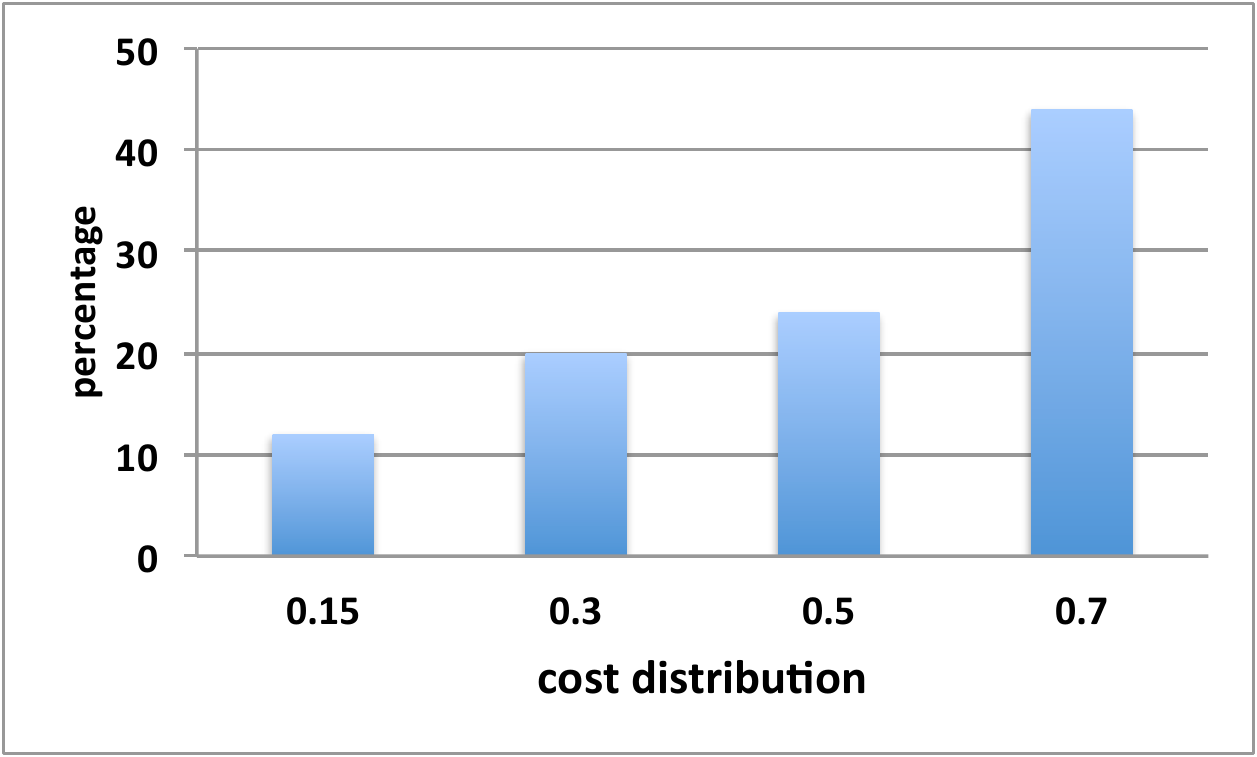}
                \caption{Worker wage distribution}
                \label{fig:scn2}
        \end{subfigure}
        \begin{subfigure}[b]{0.24\textwidth}
                \includegraphics[width=\textwidth]{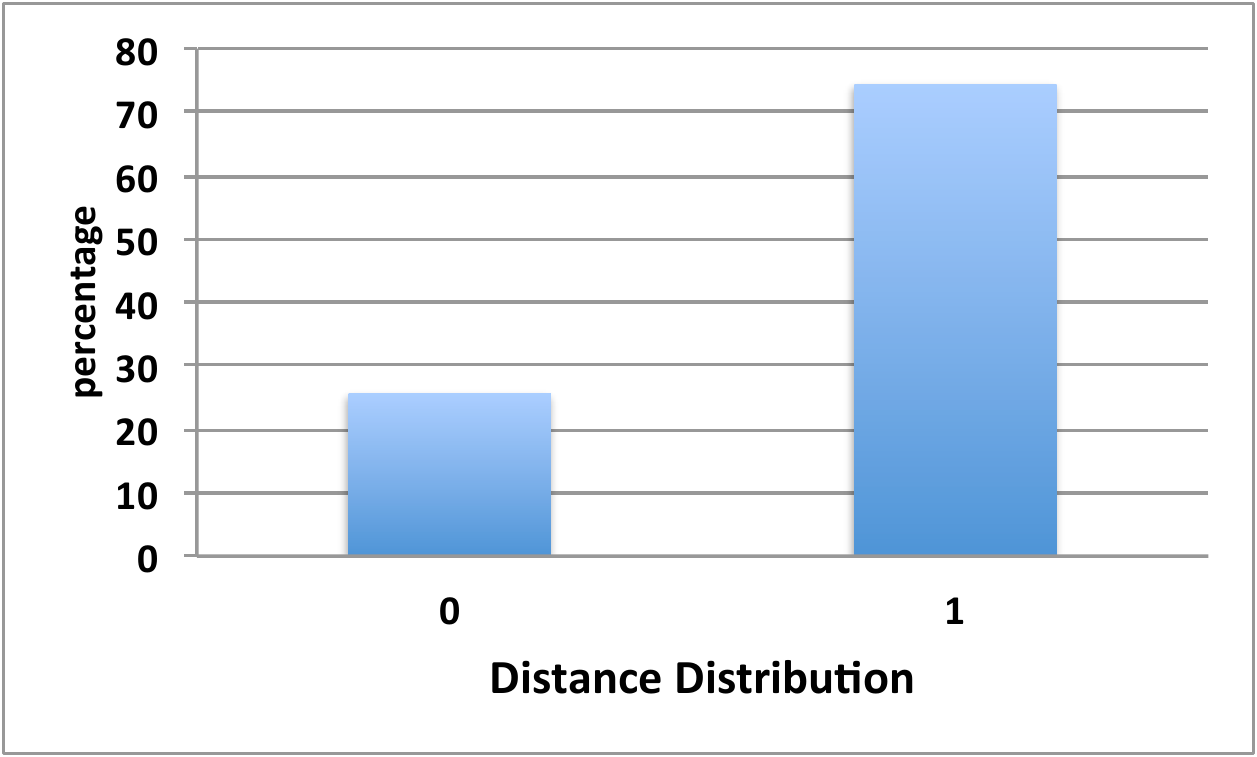}
                \caption{Worker distance distribution}
                \label{fig:scn3}
        \end{subfigure}
         \begin{subfigure}[b]{0.24\textwidth}
                \includegraphics[width=\textwidth]{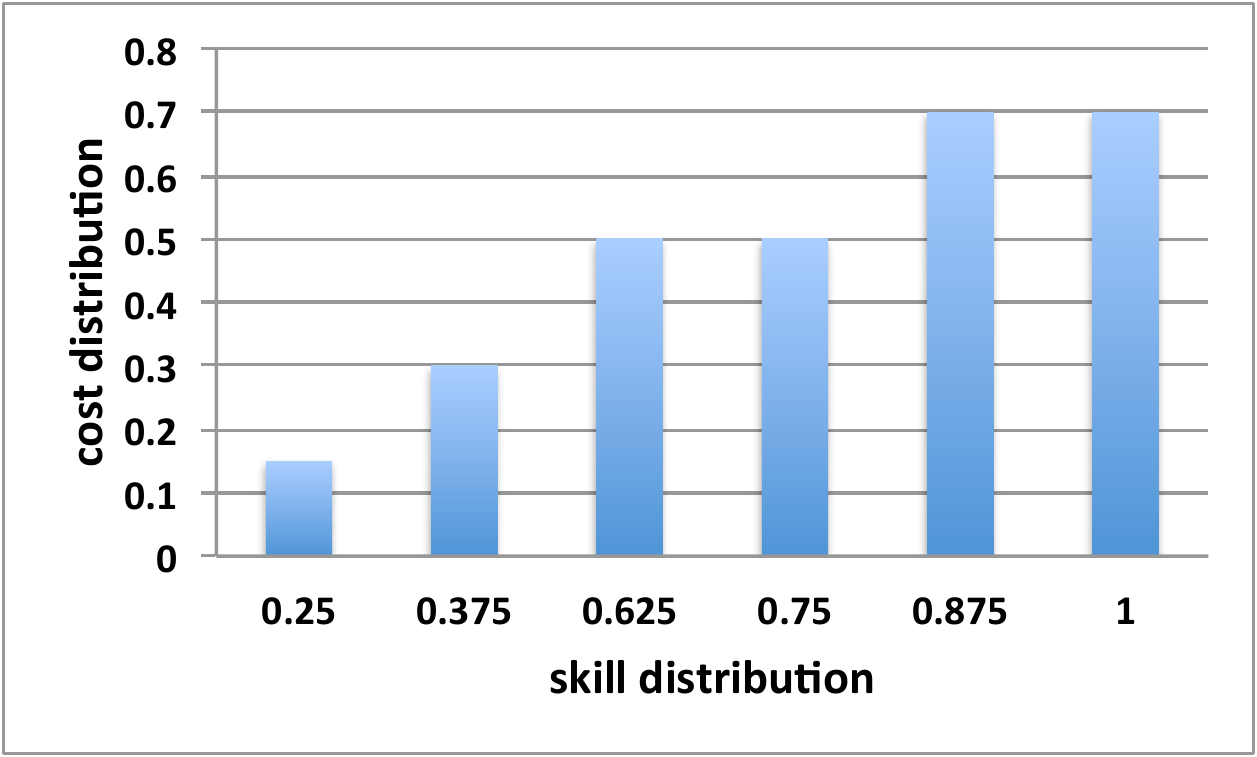}
                \caption{Strong positive correlation between worker skill and wage}
                \label{fig:scn4}
        \end{subfigure}
       \caption{Worker profile distributions for the Collaborative Document Writing in Section~\ref{sec:us}}\label{fig:phase1}
\end{figure*}

\begin{table*}[t]
\begin{tabular}{ |l|l|l|l|l|l|l|l| }
\hline
\multicolumn{8}{ |c| }{Average Rating} \\
\hline
Task & Algorithm & Completeness & Grammar & Neutrality & Clarity & Timeliness & Added-value \\ \hline
\multirow{3}{*}{MOOCs} & {\tt Optimal-CDW} & 4.6 & 4.5 & 4.3 & 4.3 & 4.3 & 3.7 \\
& {\tt Aff-Unaware-CDW} & 4.1 & 4.2 & 4.2 & 3.9 & 3.9 & 3.0\\
& {\tt CrtMass-Optimal-$10$} & 4.0 & 4.1 & 4.2 & 3.9 & 3.9 & 3.5 \\
\hline
\multirow{3}{*}{Smartphone} & {\tt Optimal} & 4.8 & 4.6 & 4.7 & 4.1 & 4.2 & 4.2\\
& {\tt Aff-Unaware} & 4.1 & 4.1 & 4.2 & 4.2 & 3.9 & 3.3\\
& {\tt CrtMass-Optimal-$10$} & 4.0 & 3.9 & 3.8 & 4.1 & 3.9 & 3.3 \\
\hline
\multirow{3}{*}{Top-10 places} & {\tt Optimal} & 4.4 & 4.2  & 4.3 & 4.2 & 4.3 & 4.3 \\
& {\tt Aff-Unaware} & 3.9 & 3.8 & 3.7 & 3.6 & 3.3 & 2.9\\
& {\tt CrtMass-Optimal-$10$} & 3.9 & 4.0 & 4.1 & 4.0 & 3.9 & 3.9 \\
\hline
\end{tabular}
\caption{\small {\bf Stage 3 results of document writing application in Section~\ref{sec:us}:} Quality assessment on the completed tasks of Stage-2 is performed by a new set of $60$ AMT workers on a scale of $1-5$. For all three tasks, the results clearly demonstrate that effective collaboration leads to better task quality. Even though all three groups (assigned to the same task) surpass the skill threhsold and satisfy the wage limit, however, our proposed formalism {\tt Optimal} enables better team collaboration, resulting in higher quality of articles.}\label{tab:userstudy}
\end{table*}

\end{document}